\documentclass[aps,prd,reprint, amsmath,amssymb,balancelastpage,nofootinbib]{revtex4-2}

\usepackage{bbold}
\usepackage{comment}
\usepackage{tensor}
\usepackage{mathtools}
\usepackage{xcolor}

\usepackage{graphicx}
\usepackage{bm}
\usepackage{hyperref}

\begin{document}

\title{The physical meaning of the Belinfante-Rosenfeld ambiguity}
\author{Ioannis Matthaiakakis}\email{i.matthaiakakis@soton.ac.uk}
\affiliation{Mathematical Sciences and STAG Research Centre, University of Southampton, Highfield,
Southampton SO17 1BJ, United Kingdom
}
\begin{abstract}
    Current literature lacks consensus on how to theoretically describe spin polarization and its transport in matter. The underlying reason is the presence of the Belinfante-Rosenfeld (BR) ambiguity in the form of the energy-momentum and spin tensors. I re-examine the BR ambiguity in Einstein-Cartan spacetimes, using the bi-form formalism in a manner applicable to any conserved current and its corresponding gauge field.\,This allows us to physically interpret the BR ambiguity as the choice of splitting a conserved current into its matter and gauge field contributions. We argue this follows from Noether's second theorem and show that certain examples of BR ambiguities reduce to large gauge transformations.
\end{abstract}

\maketitle

\textit{Introduction.} Experiments at RHIC have shown that the quark gluon plasma generated in heavy-ion collisions behaves as a nearly perfect fluid with a large angular momentum \cite{PhysRevC.98.014910,PhysRevLett.123.132301,PhysRevC.104.L061901,PhysRevC.108.014910}. The experimental results, however, cannot be fully described by standard hydrodynamics, which takes into account only the orbital part of the fluid angular momentum (vorticity). This discrepancy brought about a surge in papers analyzing the contribution of spin angular momentum to both the thermodynamics and hydrodynamics of matter \cite{Montenegro_2017,PhysRevC.97.041901,BECATTINI2019419,Florkowski_2019,PhysRevD.102.036007,Gallegos2020,Garbiso2020,10.21468/SciPostPhys.11.2.041,Becattini_2021,Hongo2021,Das_2021,Floerchinger_2022,Weickgenannt_2022,Buzzegoli_2022,Gallegos_2023, dey2024energymomentumcorrelatorsfermionsfinite,Drogosz_2025,Daher:2025pfq,Becattini:2025oyi,Zhang:2024mhs,Cartwright2025}. Despite this, no consensus exists in the literature on what spin hydrodynamics is or whether it is even useful to talk about spin transport. 

The underlying reason is an ambiguity in the local definition of the energy-momentum and angular momentum of matter. Namely, we are free to shift the energy-momentum, $T_{\mu\nu}$, and spin, $S_{\mu\nu\rho}$, tensors without affecting the conservation equations they satisfy as\footnote{We have $\Phi_{\lambda\mu\nu} = -\Phi_{\mu\lambda\nu}$, $\phi_{\lambda\mu\nu\rho} = -\phi_{\mu\lambda\nu\rho} = -\phi_{\lambda\mu\rho\nu}$ and $\Phi_{\mu[\nu\rho]} = (\Phi_{\mu\nu\rho}-\Phi_{\mu\rho\nu})/2$.} 
\begin{equation}
\label{EqIntro:BR}
    \delta T_{\mu\nu} = \partial^\lambda\Phi_{\lambda\mu\nu} ~,~\delta S_{\mu\nu\rho} = 2\Phi_{\mu[\nu\rho]} + \partial^\lambda\phi_{\lambda\mu\nu\rho}~.
\end{equation}
Equation \eqref{EqIntro:BR} is the Belinfante-Rosenfeld transformation (BRt) and the choice of particular superpotentials, $\Phi,~ \phi$, is the Belinfante-Rosenfeld ambiguity \cite{BELINFANTE1940449,Rosenfeld1940} (see also \cite{Blaschke_2016,Speranza_2021}). 

Recent examinations of the BR ambiguity have focused on the covariance properties of thermodynamics with spin under \eqref{EqIntro:BR} \cite{becattini2025pseudogauge,armas2026} (see also \cite{Jeon_2023}) or on providing an origin for certain choices of superpotentials \cite{PhysRevLett.133.262301,Iosifidis_2025}. In this paper, I re-examine the BRt \eqref{EqIntro:BR} from a cohomological point of view and show it is exact with respect to a suitably defined differential operator. This allows us to derive and extend the BRt in flat Minkowski as well as curved and torsionfull Einstein-Cartan (EC) spacetimes. I argue these extensions lead to a physical interpretation of the superpotentials and the BRt: The BRt shifts the local energy-momentum and angular momentum of matter to the background gravitational field. Thus, to use an ${\rm E\& M}$ analogy, different $\Phi,\phi$ correspond to different matter ``polarizations". This discussion generalizes to any conserved current and the gauge field it couples to.

\textit{Bi-forms.} In the following I briefly explain the bi-form formalism \cite{Labastida:1986gy,Dubois-Violette:2000fok,Dubois-Violette:2001wjr,deMedeiros:2002qpr,Bekaert:2002dt,Bruce_2019,Chatzistavrakidis_2020,Hinterbichler:2022agn,Hull:2024bcl}, employed in the sequel. 

Bi-forms are tensors which carry two sets of indices, each set taking its values in a different vector space; the $L$ and $R$ space respectively. Bi-forms are completely anti-symmetric in each set of indices, and a generic $(p,q)$ bi-form $W$ reads in components 
\begin{equation}
    \label{EqMink:GenericBiform}
    W = {1\over p!q!} W_{\mu_1\mu_2\dots\mu_p|\nu_1\nu_2\dots\nu_q} dx^{\lbrace\mu_p\rbrace}\otimes dy^{\lbrace \nu_q\rbrace}~,
\end{equation}
where $\mu_1,\dots,\mu_p$ and $\nu_1,\dots,\nu_q$ take their values in $L$ and $R$, respectively, and range from $0$ to $n-1$; $n$ being the spacetime dimension. The Einstein summation convention is used for repeated upper and lower indices, $\otimes$ stands for the tensor product between vector spaces, and 
\begin{equation}
dx^{\lbrace\mu_p\rbrace} \equiv dx^{\mu_1}\wedge dx^\mu_2\wedge\dots\wedge dx^{\mu_p} ~,
\end{equation}
is a local basis of $p$ forms in the $L$ space built out of the wedge product $\wedge$ between forms. Similarly for $dy^{\lbrace \nu_q\rbrace}$.\footnote{I should emphasize that bi-forms are tensors at a \textit{single} point, say $r$, in spacetime. However, I use $L$ and $R$ co-ordinate systems to express $r$'s co-ordinates, $x^\mu(r)$ and $y^\nu(r)$.} The vertical bar separates the $L$ and $R$ indices. 

Since bi-forms are forms in each space, we can employ all standard operators acting on forms. The most relevant for us are the $L$-exterior derivative, the $L$-Hodge star and the $L$-interior product. The $L$-exterior derivative $d$ increases the $L$-degree of a form by one via\footnote{Henceforth I suppress the wedge and tensor products.}
\begin{equation}
\label{EqMink:LExterior}
    d W = {1\over p!q!}\partial^x_{[\mu}W_{\lbrace\mu_p\rbrace]|\lbrace\nu_q\rbrace} dx^{\mu\lbrace\mu_p\rbrace} dy^{\lbrace \nu_q\rbrace}~,
\end{equation}
where the brackets denote anti-symmetrization of indices as in \eqref{EqIntro:BR}. Most importantly $d$ is nilpotent, i.e. $d^2=0$, because we antisymmetrize over commuting derivatives.

\noindent We further define the $L$-Hodge star, $\star$, as usual via
\begin{equation}
\label{EqMink:LHodge}
\star W = {1\over p!q!(n-p)!} \epsilon_{\lbrace \mu_p\rbrace\lbrace\mu_{n-p}\rbrace}W\indices{^{\lbrace \mu_p\rbrace}_{|\lbrace\nu_q\rbrace}} dx^{\lbrace\mu_{n-p}\rbrace}\otimes dy^{\lbrace\nu_q\rbrace}~, 
\end{equation}
where $\epsilon_{{\mu_1}{\mu_2}\dots{\mu_n}}$ is the completely anti-symmetric Levi-Civita symbol, satisfying $\epsilon_{012\dots n} = -1$. It is contracted to the $L$ indices of $W$ via the Minkowski metric. A useful property of $\star$ is $\star^2 W =(-1)^{p(n-p)+1}W $, i.e. it squares to the identity operator up to a sign.

\noindent Finally, the $L$-interior product operator $\iota_v$ is defined with respect to an $L$-vector $v$ and acts on a bi-form with a contraction
\begin{equation}
    \iota_v W = {1\over (p-1)!q!}v^\mu W_{\mu\lbrace\mu^{p-1}\rbrace|\lbrace\nu_q\rbrace}dx^{\lbrace\mu_{p-1}\rbrace} dy^{\lbrace \nu_q\rbrace}~.
\end{equation}

We may also define operators acting on both $L$ and $R$ indices.\,The most relevant for us is the $R$ anti-symmetrizer $\varepsilon_R$.\, It acts on a $(p,q)$ biform $W$ as
\begin{align}
    \label{EqMink:RAntisym}
   \varepsilon_R(W) ={(-1)^{q+1}\over (p-1)!q!}W_{\lbrace\mu_{p-1}\rbrace|[\mu_p\lbrace\nu_q\rbrace]} dx^{\lbrace\mu_{p-1}\rbrace} dy^{\mu_p\lbrace\nu_q\rbrace}.
\end{align} 

\noindent The $R$ anti-symmetrizer essentially turns an $L$ index of $W$ into an $R$ one. The transmutation of indices is performed via the $(1,1)$ biform $\eta  = \eta_{\mu|\nu}dx^\mu dy^\nu$ by contracting the $L$ index of $\eta$ with an $L$ index of $W$, using the Minkowski metric on $L$. We may interpret $\eta$ also as a metric when $L= R$, but in general it is more akin to a vielbein field \cite{Freedman:2012zz} as explained in a later section.

\textit{Belinfante-Rosenfeld transform in Minkowski space.}\! The BRt \eqref{EqIntro:BR} is a symmetry transformation of the relativistic energy-momentum and angular momentum conservation equations. In the present section, I rederive and extend the BRt through the use of bi-forms.

The energy-momentum and angular momentum conservation equations in Minkowski spacetime are 
\begin{equation}
\label{EqMink:ConIndex}
    \partial^\mu T_{\mu\nu} =0~,~-{1\over 2}\partial^\mu S_{\mu\nu\rho} = T_{[\nu\rho]}~.
\end{equation}
Note that the energy-momentum tensor (EMT) has no symmetry under interchanges of its indices, while the spin tensor is anti-symmetric only under the interchange of its last two indices; $S_{\mu\nu\rho} = S_{\mu[\nu\rho]}.$

To express the conservation equations \eqref{EqMink:ConIndex} in terms of bi-forms, first define the EMT and spin bi-forms 
\begin{equation}
\label{EqMink:TSBiforms}
    T = T_{\mu|\nu}dx^\mu dy^\nu ~,~S = {1\over 2}S_{\mu|\nu_1\nu_2} dx^\mu dy^{\nu_1\nu_2}~.
\end{equation}
The definition \eqref{EqMink:TSBiforms} is mostly motivated by the symmetry properties of $T_{\mu\nu}$ and $S_{\mu\nu\rho}$.\,Further motivation stems from the 1st order and metric-affine gravity formalisms \cite{Freedman:2012zz, Hehl:1994ue,jiménezcano2022metricaffinegaugetheoriesgravity}, where $T$ and $S$ naturally take their values in two vector spaces, precisely as in \eqref{EqMink:TSBiforms}.

\noindent In terms of $T$, $S$, the conservation equations \eqref{EqMink:ConIndex} become
\begin{equation}
\label{EqMink:ConBi}
    \star d\star T = 0 ~,~\star d \star  S = \varepsilon_R(T)~,
\end{equation}
as can be verified by using standard manipulations of  \eqref{EqMink:LExterior}, \eqref{EqMink:LHodge} (see e.g. \cite{Carroll_2019}), and \eqref{EqMink:RAntisym}.

\noindent Equation \eqref{EqMink:ConBi} suggests the natural variables to work with are $\star T$ and $\star S$. So, define 
\begin{equation}
    B^1 = \star T~,~B^2 = \star S, 
\end{equation}
and write \eqref{EqMink:ConBi} as
\begin{equation}
\label{EqMink:ConBi2}
    dB^1 = 0~,~dB^2 = \eta B^1~.
\end{equation}
The product between $\eta$ and $B^1$ is the wedge product between forms in both $L$ and $R$ spaces. 

In the form \eqref{EqMink:ConBi2}, the nilpotency of $d$, $d^2=0$, implies two important corollaries. First, we find 
\begin{equation}
\label{EqMink:VanishingT}
    \tau \equiv d\eta = 0~,
\end{equation}
by acting with $d$ on \eqref{EqMink:ConBi2}. Thus $\eta$ must be irrotational. 

\noindent Second, it is clear that
\begin{equation}
\label{EqMink:BR1}
    \delta B^1 = d\Xi^1~,~\delta B^2 = d\Xi^2 - \eta\Xi^1~,
\end{equation}
is a symmetry of \eqref{EqMink:ConBi2}.\,Equation \eqref{EqMink:BR1} is a re-write of the BRt \eqref{EqIntro:BR} with $\Xi^1$ and $\Xi^2$ the $\star$ of $\Phi$ and $\phi$, respectively. 

Thus, expressing the conservation equations in terms of bi-forms makes manifest the BRt, due to $d^2=0$. However we gain even more from bi-forms: The constraint \eqref{EqMink:VanishingT} is invariant under $ \delta\eta = d\xi~,$ with $\xi$ a $(0,1)$ bi-form. To interpret this invariance, note that \eqref{EqMink:VanishingT} locally implies $\eta = dx_R$ with $x_R$ the position co-vector in $R$.\footnote{In index notation $\eta_{\mu\nu}= \partial_\mu x_\nu$.} Hence, the shift of $\eta$ is induced by a translation of $x_R$; $\delta x_R = \xi$.
Consequently, we must interpret $\eta$ as a gauge field of translations,\footnote{We cannot identify $\eta$ as a vielbein in Minkowski spacetime, because of this gauge field interpretation.} which implies $\tau$ is geometrically the torsion of spacetime \cite{Hehl:1994ue}. 

More importantly, $R-$translations by $\xi$ can be extended to a symmetry of the conservation equations \eqref{EqMink:ConBi2}. In particular, \eqref{EqMink:ConBi2} is invariant under $R-$translations if
\begin{equation}
\label{EqMink:SpinTranslation}
   \delta_\xi\eta = d\xi~,~ \delta_\xi B^2 = \xi B^1~,~\delta_\xi B^1 =0. 
\end{equation}
Translations by $\xi$ mix $B^1$ and $B^2$, implying we should think of them as the doublet $B^I = (B^1,B^2)$. In terms of $B^I$, the conservation equations \eqref{EqMink:ConBi2} become a single vectorial equation
\begin{equation}
\label{EqMink:ConBi3}
    dB^I-\eta C^I_JB^J \equiv DB^I \equiv {\cal F}^I = 0~.
\end{equation}
In \eqref{EqMink:ConBi3}, I first introduced the matrix $C^I_J = \delta^I_2\delta^1_J$, then the covariant derivative operator $D = d\delta^I_J -\eta C^I_J$ acting on $B^I$. Finally, I defined ${\cal F}^I = DB^I$ for later convenience. 

\noindent Taking the square of $D$ we find 
\begin{equation}
    D^2B^I = -\tau C^I_JB^J~,
\end{equation}
implying $D$ is the exterior covariant derivative of translations, and $B^I$ transforms under the $R-$translation representation generated by $C^I_J$. 

\noindent The introduction of $\xi$ then extends the BRt \eqref{EqMink:BR1} to
\begin{equation}
\label{EqMink:BRtDoublet}
    \delta \eta = d\xi ~,~ \delta B^I = D\Xi^I +\xi C^I_JB^J~.
\end{equation}
See App.~\ref{AppB:EMJExamples} on how to use \eqref{EqMink:BRtDoublet} to fix the form of $B^I$.

Equation \eqref{EqMink:BRtDoublet} is of course also a symmetry of ${\cal F}^I$ since
\begin{equation}
    \label{EqMink:BRtF}
    \delta {\cal F}^I = \xi C^I_J{\cal F}^J -\tau C^I_J\Xi^J= 0~.
\end{equation}
As a comment, note that the conservation equations are invariant under the BRt \eqref{EqMink:BRtDoublet} only when the microscopic equations of motion hold, i.e. when the theory is \textit{on-shell}. This implies that \eqref{EqMink:BRtDoublet} is a symmetry of the solution space of a microscopic theory, and does not necessarily follow from a symmetry of the microscopic Lagrangian. This is also suggested in \cite{HEHL197655}.

The physical meaning of \eqref{EqMink:BRtDoublet} can be gleaned from its action on the globally conserved charges. These are the total energy-momentum and angular momentum
\begin{equation}
    P = \int_{\Sigma_L} B^1 ~,~ J = \int_{\Sigma_L} x_R B^1 - B^2~,
\end{equation}
where the integral is taken over a co-dimension one $L$-surface. The BRt \eqref{EqMink:BRtDoublet} acts on $P$ and $J$ via
\begin{equation}
\label{EqMinkBRtGlobal}
    \delta P = \int_{S_L} \Xi^1 ~,~ \delta J = \int_{S_L} x_R \Xi^1 - \Xi^2~,
\end{equation}
which tells us the BRt \eqref{EqMink:BR1} shifts the total energy-momentum and angular momentum by the fluxes of the superpotentials through a codimension two surface $S_L$.  In contrast, $R$-translations leave the charges invariant. The latter fact clarifies the physical origin of the transformation of $B^2$ under $R$-translations: $B^2$ must transform under $R$-translations in order to compensate for the orbital angular momentum added to the system due to the shift of the spacetime origin.

Finally, let us explore the group structure of \eqref{EqMink:BRtDoublet} by calculating the algebra of successive transformations. Consider the transformations with parameters $(\xi_1, \Xi^I_1)$ and $(\xi_2, \Xi^I_2)$, respectively. Calculating the commutator of these two transformations on $B^I$, we find it closes to a third BRt, i.e. 
\begin{equation}
\label{EqMink:Commutator}
    [\delta_1,\delta_2]B^I = \delta_3B^I ~,
\end{equation}
with $\delta_3$ parametrized by (see App.\ref{AppA})
\begin{equation}
\label{EqMink:ThirdT}
    (\xi_3, \Xi^I_3) = (0, \xi_2C^I_J\Xi^J_1- \xi_1C^I_J\Xi^J_2)~.
\end{equation}
We see that separately the $\xi$s and the $\Xi$s form abelian subgroups, since no $\xi^2$ or $(\Xi^I)^2$ terms appear in $\delta_3$. However, the $R-$translations have a non-trivial action on the superpotentials. Therefore, the full BRt group preserving the conservation equations in Minkowski spacetime is the semi-direct product
\begin{equation}
\label{EqMink:BRtGroup}
    G_{\rm M-BR} \equiv R^n\ltimes \left(R^n \times R^{n(n-1)/2)}\right) = R^n\ltimes R^{n(n+1)/2)}~,
\end{equation}
where the left factor denotes the group of $R$-translations and the right factor the shift of the energy-momentum and angular momentum of the system. The $\ltimes$ product indicates that the left factor acts non-trivially on the right one via \eqref{EqMink:ThirdT}.

\noindent Equation \eqref{EqMink:BRtGroup} constitutes an extension of the usually discussed BRt group and, along with its action \eqref{EqMink:BRtDoublet} on $\eta$ and $B^I$, constitutes one of the main results of this paper. 

\textit{Extension to Einstein-Cartan spacetime.} The $G_{\rm M-BR}$ group in Eq.~\eqref{EqMink:BRtGroup} showcases the importance of local $R$-symmetries in the BRt discussion. It also suggests that extending the present analysis to more general spacetimes with non-zero torsion $\tau$ might modify our discussion. For these reasons, in this section I derive the BR group in an EC spacetime, i.e. a spacetime with non-trivial torsion and curvature \cite{Trautman:2006fp}. 

From this point forward, I interpret $L$ as the local tangent space over a point in our spacetime and $R$ as a local Minkowski space above the same point. This allows us to map the bi-form formalism to the standard formulation of gravity in an EC spacetime \cite{Hehl:1994ue,Trautman:2006fp, jiménezcano2022metricaffinegaugetheoriesgravity}. Namely, we can identify $\eta$ with the vielbein field in said formulations. The only issue preventing this matching is $\eta$'s non-trivial transformation under $R$-translations, \eqref{EqMink:BRtDoublet}. This can be remedied by examining the conservation equations. 

The conservation equations can be read-off of \cite{Hehl:1994ue,jiménezcano2022metricaffinegaugetheoriesgravity} by restricting to the EC limit. We have\footnote{See \cite{jiménezcano2022metricaffinegaugetheoriesgravity} for co-ordinate expressions.}
\begin{equation}
\label{EqEC:Con1}
    \nabla B^1 = (\iota_\eta \tau)\cdot_R B^1 - (\iota_\eta \Omega)\cdot_R B^2~,~\nabla B^2 = 0~.
\end{equation}
Several new notations were introduced in \eqref{EqEC:Con1}. First, the exterior covariant derivative $\nabla$ is defined as
\begin{equation}
    \nabla = D + [\omega, ] = \overset{\omega}{\nabla} -\eta ~.
\end{equation}
$\omega$ is the spin connection with components $(\omega_\mu)\indices{^a_b}$ and $[\omega, \cdot]$ is an $L-$graded commutator defined as \cite{Bertlmann:1996xk},
\begin{equation}
\label{EqEC:GradedCom}
    [\omega, W] = \omega\cdot W +(-1)^{p+1} W\cdot \omega~.
\end{equation}
I denote by $\cdot$ the matrix product in $R$. For example, 
\begin{equation}
    (W\cdot\omega)^a_{~bc} = W_{bd}\wedge \omega^d_{~c} + W_{dc}\wedge\omega^{c}_{~b}~,~ \omega\cdot W = 0
\end{equation}
for a $(p,2)$ bi-form $W$, and analogously for general $R$ tensors.

\noindent The $(2,1)$ and $(2,2)$ biforms $\tau$ and $\Omega$ are the torsion and curvature of the EC spacetime defined via
\begin{align}
\label{EqEC:NablaSquare}
    &\nabla^2 B^I = -\tau C^I_JB^J + [\Omega, B^J]~,
    \\
    &\tau = \nabla \eta~,~\Omega = d\omega + {1\over 2} [\omega, \omega]~.
\end{align}
Further, in \eqref{EqEC:Con1} I treat $\eta$ as an $L-$vector by raising its $L$-index with a general $L$-metric $g$. Thus the $\iota_\eta$ contraction changes an $L$-index to an $R$-index (without antisymmetrization). For example, $(\iota_\eta\tau)_{\mu a|b} = g^{\nu\rho}\eta_{\nu|a}\tau_{\rho\mu|b}$. The inner product $\cdot_R$ contracts the $R$ indices of two bi-forms using the Minkowski metric.\footnote{$R-$indices due to $\iota_\eta$ are not contracted.} For example, $\tau\cdot_R B^1= \tau^a\wedge B^{1}_a$ with $L$ indices suppressed. Finally, recall that only $B^2$ is charged under translations, such that $\nabla B^1 = \overset{\omega}{\nabla}B^1$ and $\nabla B^2 = \overset{\omega}{\nabla}B^2 -\eta B^1$.

The BRt group in EC spacetimes is thus the group of transformations preserving \eqref{EqEC:Con1}. We suggest the infinitesimal form of the BRt is 
\begin{align}
    &\delta\eta = \eta\cdot\theta + \nabla \xi ~,~\delta\omega = \nabla\theta~,~\delta B^2 = B^2\cdot\theta+ \nabla\Xi^2 + \xi B^1~, \nonumber
    \\
    \label{EqEC:BR1}
    \\
    &\delta B^1= B^1\cdot\theta +\nabla\Xi^1 -(\iota_\eta\tau)\cdot_R\Xi^1 +(\iota_\eta\Omega)\cdot_R\Xi^2~.\nonumber
\end{align}
The newly introduced matrix $\theta^a_{~b}$ parameterizes local $R$-rotations and the $\cdot_R$ product follows the same rules as explained below \eqref{EqEC:GradedCom}. The resulting transformation for ${\cal F}^I = \nabla B^I$ is also modified to
\begin{equation}
    \label{EqEC:ConBR1}
    \delta{\cal F}^I = {\cal F}^I\cdot\theta + \xi C^I_J {\cal F}^J~.
\end{equation}
Importantly, notice that when $\xi \neq 0$ the angular momentum conservation equation is shifted by a term proportional to the left-hand side of the energy-momentum conservation equation. This means that upon enforcing \eqref{EqEC:Con1}, the angular momentum conservation equation is not satisfied in general. In short, we have found that $R$-translations are no longer a symmetry of our conservation equations and they must be frozen. This is natural from the point of view of differential geometry, since there is no canonical map that would allow us to associate two vector spaces at two different points in spacetime.

Therefore, the true BRt of \eqref{EqEC:Con1} is
\begin{align}
    &\delta\eta = \eta\cdot\theta ~,~\delta\omega = \nabla\theta~,~\delta B^2 = B^2\cdot\theta+ \nabla\Xi^2~,~  \nonumber
    \\
    \label{EqEC:BR2}
    \\
    &\delta B^1= B^1\cdot\theta +\nabla\Xi^1 -(\iota_\eta\tau)\cdot_R\Xi^1 +(\iota_\eta\Omega)\cdot_R\Xi^2~.\nonumber
\end{align}
Essentially, \eqref{EqEC:BR2} tells us that $\eta$ is in fact the vielbein field between the local tangent space and Minkowski spacetime \cite{Hehl:1994ue}.

\noindent In index notation, the BRt \eqref{EqEC:BR2} reads at $\theta =0$
\begin{widetext}
\begin{align}
\label{EqEC:BelTindex}
    &\delta T_{\mu |a} ={1\over \sqrt{-g}}\overset{\omega}{\nabla}_\nu\left(\sqrt{-g} \Phi\indices{^\nu_{\mu|a}}\right) +\tau\indices{_{\nu a}^{|b}} \Phi\indices{^\nu_{\mu |b}} -\Omega\indices{_{\nu a}^{|c_1 c_2}} \phi\indices{^\nu_{\mu |c_1c_2}}~,
    \\
    \label{EqEC:BelSindex}
    &\delta S_{\mu|ab} = {1\over \sqrt{-g}}\overset{\omega}{\nabla}_\nu\left(\sqrt{-g} \phi\indices{^\nu_{\mu|ab}}\right)  + 2\Phi_{\mu|[ab]} ~.
\end{align}
\end{widetext}
with $g$ the $L-$metric determinant, $\Phi = (-1)^n \star\Xi^1$, $\phi = (-1)^n \star\Xi^2$, $\tau\indices{_{\mu a|b}} = (\iota_\eta\tau)_{\mu a|b}$ and similarly for $\Omega\indices{_{\mu a |b c}}$. 

\noindent The EC-BR group  corresponding to \eqref{EqEC:BR2} is (see App. \ref{AppA})
\begin{equation}
\label{EqEC:BRtGroup}
 G_{\rm EC-BR} = SO(n-1,1)\ltimes R^{n(n+1)/2}~.   
\end{equation}

Finally, I should note that most of \eqref{EqEC:BR2} appears already in \cite{Hehl:1994ue}. However, the curvature term in \eqref{EqEC:BelTindex} is new as far as I am aware. It is precisely this new term that allows us to give a concrete physical meaning to the BRt parameters.

\textit{The physical meaning of the superpotentials.} We have now all the necessary tools to understand the physical meaning of the superpotentials $\Xi^I$. To do so, consider the pure ``gauge" modes of $G_{\rm EC-BR}$. These are the EMT and spin tensors BR-equivalent to zero, i.e.
\begin{align}
\label{EqEC:EMPG}
    &B^1_{\rm pg} = \nabla\Xi^1 -(\iota_\eta\tau)\cdot_R\Xi^1 +(\iota_\eta\Omega)\cdot_R\Xi^2~,
    \\
    \label{EqEC:SpinPG}
    &B^2_{\rm pg} = \nabla\Xi^2 = \overset{\omega}{\nabla}\Xi^2 - \eta\Xi^1~,
\end{align}
The most important feature of equations \eqref{EqEC:EMPG} and \eqref{EqEC:SpinPG} is that they are formally the same as the equations satisfied by a dynamical EC theory of gravity \cite{Hehl:1994ue, jiménezcano2022metricaffinegaugetheoriesgravity}.\footnote{Note our spin tensor is minus the spin tensor in \cite{Hehl:1994ue,jiménezcano2022metricaffinegaugetheoriesgravity}.} To be precise, in EC gravity we have
\begin{align}
\label{EqEC:EMGrav}
    &\nabla H^1 = B^1 + (\iota_\eta\tau)\cdot_R H^1 -(\iota_\eta\Omega)\cdot_R H^2~,~
    \\
    \label{EqEC:SpinGrav}
    &\overset{\omega}{\nabla} H^2 = B^2 + \eta H^1~.
\end{align}
where
\begin{align}
\label{EqEC:Variations}
    &H^1 = -{\partial L_{\rm EC}\over \partial\tau}~,~H^2 = {\partial L_{\rm EC}\over \partial\Omega}~,
    \\
    &B^1 = B_{\rm mat} + \iota_{\eta}L_{\rm EC}~,~ B^2 = B^2_{\rm mat}~,
\end{align}
with $L_{\rm EC}$ the gravitational Lagrangian $(n,0)$ form, and $B_{\rm mat}$ the matter EMT and similarly for $B^2_{\rm mat}$. The fields $H^1$ and $H^2$ are called the gravitational momenta of $L_{\rm EC}$.

\begin{figure}[h]
     \centering
     \includegraphics[scale=0.275]{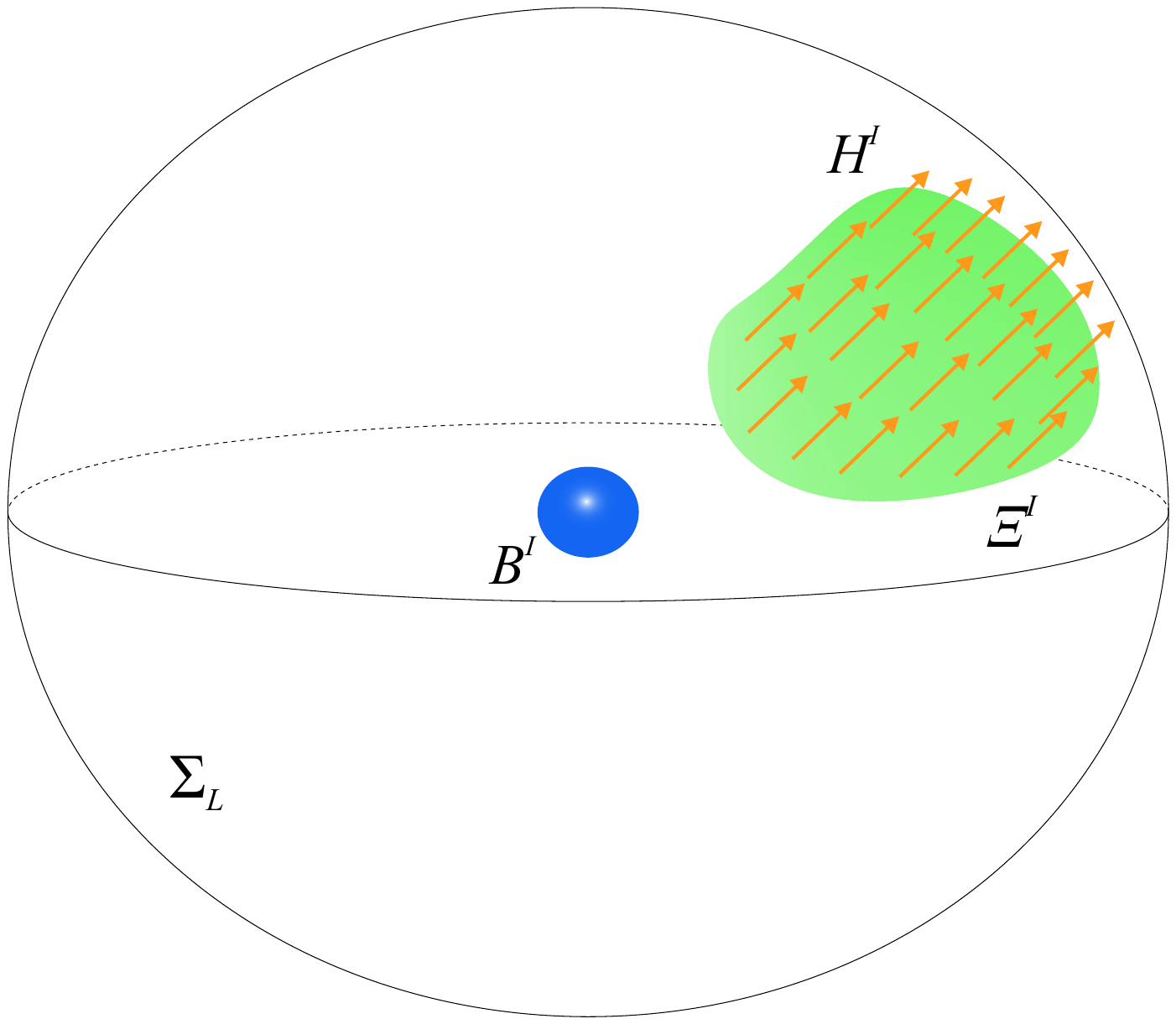}
     \caption{A conserved charge (blue sphere) $B^I$ in a volume $\Sigma_L$ can be (partially or fully) delocalized on the boundary of $\Sigma_L$ using a BRt described by $\Xi^I$ (green region). The resulting surface charge can be interpreted as the flux (orange arrows) of the gauge momenta $H^I$ through the boundary of $\Sigma_L$.  }
     \label{fig:Fluxes}
 \end{figure}

 We see then that the the BRt \eqref{EqEC:BR2} when applied to Eq.s \eqref{EqEC:EMGrav} and \eqref{EqEC:SpinGrav}, shifts parts of the energy-momentum and spin of matter to the gravitational momenta. This implies that by choosing specific superpotentials we define which part of the energy-momentum and angular momentum of matter we consider to be part of the gravitational field (see Fig. \ref{fig:Fluxes}). This is the second major result of this work.

To showcase the naturality of this interpretation, I note that the formalism I employ in this paper works for any background field; not just curvature and torsion. Thus, for example, consider E$\&$M coupled to matter. In this case, we know that the BRt amounts to shifting the polarization and magnetization tensor from the bound current and charge in matter to the E$\&$M field  \cite{landau, Jackson:1998nia,Kovtun_2016}. 

The conservation equations for a U(1) current $(n-1,0)$ form $j$ coupled to an electromagnetic field described by the Maxwell $(2,0)$ form $F$ are in bi-form notation
\begin{equation}
\label{EqEC:U1}
    dB^1 = (\iota_\eta F) j~,~ dj =0~.
\end{equation}
It is straightforward to guess the BRt preserving \eqref{EqEC:U1}, by noting the similarity between \eqref{EqEC:U1} and \eqref{EqEC:Con1}. It reads 
\begin{equation}
\label{EqEC:BRU1}
    \delta B^1 = d\Xi^1 -\iota_\eta F\Xi^2~,~ \delta j = d\Xi^2~.
\end{equation}
The pure ``gauge" modes of \eqref{EqEC:BRU1} correspond to the dynamical equations
\begin{equation}
\label{EqEC:Maxwell}
    dH^1 = B^1 +\iota_\eta F H^2 ~,~ dH^2 = j~,
\end{equation}
where
\begin{equation}
\label{EqEC:MaxVariations}
    B^1 = B_{\rm mat} +\iota_\eta L_{\rm EM}~,~ H^2 = -{\partial L_{\rm EM}\over \partial F}~,
\end{equation}
and $L_{\rm EM}$ is the electromagnetic Lagrangian $(n,0)$ form. 

\noindent The first equation of \eqref{EqEC:Maxwell} tells us that it is the EMT of both matter and the electromagnetic field, $B^1 +\iota_\eta F H^2$, that is conserved, while the second gives the Maxwell equations in matter. Thus upon using \eqref{EqEC:BRU1}, we see $\Xi^2$ shifts bound currents from $j$ to $H^2$, as expected. We can similarly extend our discussion to any conserved current. 

I believe the universality of this conclusion stems from Noether's second theorem \cite{Noether1918,Noether01011971}, which states that gauging a symmetry implies the \textit{off-shell} conservation of its corresponding matter current. As a result, the BR ambiguity in the current must also result in an ambiguity in the gauge theory it couples to (see also App.\ref{AppC:ActionBRt}). The above abstract analysis is a proof of that statement for Poincare and U(1) gauge theories.

 One may then wonder whether a BRt can be realized as a gauge field transformation. This is true when
\begin{equation}
\label{Eq:GenConditions}
\delta H^I = \Xi^I~,
\end{equation}
can be solved in terms of the corresponding gauge field. Whether this is possible relies on the precise form of the gauge field Lagrangian. In the simplest case of E$\&$M in vacuum, \eqref{Eq:GenConditions} simplifies to (see App.\ref{AppC:ActionBRt}) 
\begin{equation}
\label{Eq:Conditions}
   d \delta\eta = -\star\Xi^1~,~ d \delta A = -\star\Xi^2~,
\end{equation}
with $A$ a U(1) gauge field. We can solve for $\delta\eta$ and $\delta A$ only if $d\star\Xi^{1,2}=0$,  to find
\begin{equation}
\label{Eq:GaugeT}
    \delta\eta = \lambda^1 + d\xi~,~~\delta A = \lambda^2 + d\chi~,
\end{equation}
with $d\lambda^I = -\star\Xi^I$ and $\xi,~ \chi$ arbitrary. Equation \eqref{Eq:GaugeT} is an example of large $R$-translation and U(1) gauge transformations \cite{HenneauxTeitelboim}, consistent with the fact that a BRt may change the total energy-momentum and charge of the system. Thus gauge transformations are a subclass of BRts, at least for the particular case of vacuum E$\&$M. Whether this conclusion is true in general should be confirmed with an in-depth analysis of Eq.\eqref{Eq:GenConditions}.

\textit{Conclusions\! $\&$\! Outlook.}\! In this paper, I employed the bi-form formalism in order to understand the Belinfante-Rosenfeld transformation group of any conserved current in the presence of any background field. In particular, I applied this formalism to the case of conserved energy-momentum and angular momentum tensors in both Minkowski and Einstein-Cartan spacetimes. This has led us to an extension of the Belinfante-Rosenfeld transformation, \eqref{EqMink:BRtDoublet} and \eqref{EqEC:BR2}, respectively. These extensions were crucial in deriving the true group structure of the Belinfante-Rosenfeld transformation, \eqref{EqMink:BRtGroup} and \eqref{EqEC:BRtGroup}, as well as the physical meaning of the Belinfante-Rosenfeld transformation parameters. Namely, I have found that the superpotentials shift the conserved current from matter to the gauge field coupling to said current (see discussion around \eqref{EqEC:EMGrav}). That is, the Belinfante-Rosenfeld transformation parameters are analogous to the polarization and magnetization tensors in E$\&$M.

An important outlook of this work is in spin hydrodynamics. In particular, my results imply that comparing spin-hydrodynamics analyses performed for different choices of the superpotentials amounts to comparing systems with different ``free'' energy-momentum and angular momentum. Thus, to compare systems for different choices of superpotentials, we need to construct a Belinfante-Rosenfeld-covariant spin hydrodynamics. The authors of \cite{Montenegro2026phf} provide one such construction, by considering fluctuations of the thermal partition function around equilibrium and the Ward-identities it satisfies. We may also construct a Belinfante-Rosenfeld-covariant spin hydrodynamics by following the example of \cite{Kovtun_2016,Amoretti:2022ovc,Amoretti:2024jig} for polarized matter. Namely, we need to incorporate torsion and curvature into the thermodynamics of matter.\footnote{Part of this analysis is performed in \cite{Kovtun_2020}, where curvature effects are taken into account for spinless matter.} This is also suggested by the discussion around Eq.\eqref{Eq:GenConditions}, where $H^I$ is used to compensate a BRt. I must emphasize that this is necessary only to compare spin hydrodynamics at different superpotentials. To construct a canonical theory of spin hydrodynamics, we should simply set the superpotentials to zero. This amounts to using the canonical energy-momentum and spin tensors, as suggested in \cite{HEHL197655} more generally, and explicitly carried out in \cite{Gallegos_2023}.

I have formulated the Belinfante-Rosenfeld transformation in terms of exact forms with respect to a suitably defined operator. I have also shown it reduces to a large gauge transformation if restricted appropriately.It would then be interesting to understand if we can go further and reformulate the Belinfante-Rosenfeld symmetry as a BRST gauge symmetry \cite{Weinberg:1996kr}.\footnote{Work in this direction appeared recently in \cite{jonsson2026hydrodynamicscospansfieldtheories}, independently of this work, for the case of magnetohydrodynamics.} 

Finally, we have seen that the superpotentials imply the presence of ambiguities in the definition of the gravitational momenta. Therefore it would be interesting to understand the Belinfante-Rosenfeld ambiguity from the point of view of gravity, where similar ambiguities arise in the covariant phase space approach \cite{Lee:1990nz,Wald:1999wa,Harlow_2020}. Can the conventions used in this setting to fix the ambiguity in the conserved currents also be used to fix the energy-momentum and spin tensors?

\begin{acknowledgements}
    \textit{Acknowledgments.}\! I would like to thank Prahar Mitra and Kostas Skenderis for helpful discussion on a previous version of this work. The work of I.M. is supported by the STFC consolidated grant (ST/X000583/1) ‘New Frontiers In Particle Physics, Cosmology And Gravity’.

    AI use: Anthropic's Claude AI Sonnet 4.6 was used to proofread a draft of the document and to simulate two rounds of peer review. All calculations and writing in this paper were performed by IM.
\end{acknowledgements}

\bibliography{biblio}

\begin{thebibliography}{66}%
\makeatletter
\providecommand \@ifxundefined [1]{%
 \@ifx{#1\undefined}
}%
\providecommand \@ifnum [1]{%
 \ifnum #1\expandafter \@firstoftwo
 \else \expandafter \@secondoftwo
 \fi
}%
\providecommand \@ifx [1]{%
 \ifx #1\expandafter \@firstoftwo
 \else \expandafter \@secondoftwo
 \fi
}%
\providecommand \natexlab [1]{#1}%
\providecommand \enquote  [1]{``#1''}%
\providecommand \bibnamefont  [1]{#1}%
\providecommand \bibfnamefont [1]{#1}%
\providecommand \citenamefont [1]{#1}%
\providecommand \href@noop [0]{\@secondoftwo}%
\providecommand \href [0]{\begingroup \@sanitize@url \@href}%
\providecommand \@href[1]{\@@startlink{#1}\@@href}%
\providecommand \@@href[1]{\endgroup#1\@@endlink}%
\providecommand \@sanitize@url [0]{\catcode `\\12\catcode `\$12\catcode `\&12\catcode `\#12\catcode `\^12\catcode `\_12\catcode `\%12\relax}%
\providecommand \@@startlink[1]{}%
\providecommand \@@endlink[0]{}%
\providecommand \url  [0]{\begingroup\@sanitize@url \@url }%
\providecommand \@url [1]{\endgroup\@href {#1}{\urlprefix }}%
\providecommand \urlprefix  [0]{URL }%
\providecommand \Eprint [0]{\href }%
\providecommand \doibase [0]{https://doi.org/}%
\providecommand \selectlanguage [0]{\@gobble}%
\providecommand \bibinfo  [0]{\@secondoftwo}%
\providecommand \bibfield  [0]{\@secondoftwo}%
\providecommand \translation [1]{[#1]}%
\providecommand \BibitemOpen [0]{}%
\providecommand \bibitemStop [0]{}%
\providecommand \bibitemNoStop [0]{.\EOS\space}%
\providecommand \EOS [0]{\spacefactor3000\relax}%
\providecommand \BibitemShut  [1]{\csname bibitem#1\endcsname}%
\let\auto@bib@innerbib\@empty
\bibitem [{\citenamefont {Adam}\ \emph {et~al.}(2018)\citenamefont {Adam}, \citenamefont {Adamczyk}, \citenamefont {Adams}, \citenamefont {Adkins}, \citenamefont {Agakishiev}, \citenamefont {Aggarwal}, \citenamefont {Ahammed} \emph {et~al.}}]{PhysRevC.98.014910}%
  \BibitemOpen
  \bibfield  {author} {\bibinfo {author} {\bibfnamefont {J.}~\bibnamefont {Adam}}, \bibinfo {author} {\bibfnamefont {L.}~\bibnamefont {Adamczyk}}, \bibinfo {author} {\bibfnamefont {J.~R.}\ \bibnamefont {Adams}}, \bibinfo {author} {\bibfnamefont {J.~K.}\ \bibnamefont {Adkins}}, \bibinfo {author} {\bibfnamefont {G.}~\bibnamefont {Agakishiev}}, \bibinfo {author} {\bibfnamefont {M.~M.}\ \bibnamefont {Aggarwal}}, \bibinfo {author} {\bibfnamefont {Z.}~\bibnamefont {Ahammed}}, \emph {et~al.} (\bibinfo {collaboration} {STAR Collaboration}),\ }\bibfield  {title} {\bibinfo {title} {Global polarization of $\mathrm{\ensuremath{\Lambda}}$ hyperons in au + au collisions at $\sqrt{{s}_{NN}}=200$ gev},\ }\href {https://doi.org/10.1103/PhysRevC.98.014910} {\bibfield  {journal} {\bibinfo  {journal} {Phys. Rev. C}\ }\textbf {\bibinfo {volume} {98}},\ \bibinfo {pages} {014910} (\bibinfo {year} {2018})}\BibitemShut {NoStop}%
\bibitem [{\citenamefont {Adam}\ \emph {et~al.}(2019)\citenamefont {Adam}, \citenamefont {Adamczyk}, \citenamefont {Adams}, \citenamefont {Adkins}, \citenamefont {Agakishiev}, \citenamefont {Aggarwal}, \citenamefont {Ahammed} \emph {et~al.}}]{PhysRevLett.123.132301}%
  \BibitemOpen
  \bibfield  {author} {\bibinfo {author} {\bibfnamefont {J.}~\bibnamefont {Adam}}, \bibinfo {author} {\bibfnamefont {L.}~\bibnamefont {Adamczyk}}, \bibinfo {author} {\bibfnamefont {J.~R.}\ \bibnamefont {Adams}}, \bibinfo {author} {\bibfnamefont {J.~K.}\ \bibnamefont {Adkins}}, \bibinfo {author} {\bibfnamefont {G.}~\bibnamefont {Agakishiev}}, \bibinfo {author} {\bibfnamefont {M.~M.}\ \bibnamefont {Aggarwal}}, \bibinfo {author} {\bibfnamefont {Z.}~\bibnamefont {Ahammed}}, \emph {et~al.} (\bibinfo {collaboration} {STAR Collaboration}),\ }\bibfield  {title} {\bibinfo {title} {Polarization of $\mathrm{\ensuremath{\Lambda}}$ ($\overline{\mathrm{\ensuremath{\Lambda}}}$) hyperons along the beam direction in $\mathrm{Au}+\mathrm{Au}$ collisions at $\sqrt{{s}_{NN}}=200\text{ }\text{ }\mathrm{GeV}$},\ }\href {https://doi.org/10.1103/PhysRevLett.123.132301} {\bibfield  {journal} {\bibinfo  {journal} {Phys. Rev. Lett.}\ }\textbf {\bibinfo {volume} {123}},\ \bibinfo {pages} {132301} (\bibinfo {year} {2019})}\BibitemShut
  {NoStop}%
\bibitem [{\citenamefont {Abdallah}\ \emph {et~al.}(2021)\citenamefont {Abdallah}, \citenamefont {Aboona}, \citenamefont {Adam}, \citenamefont {Adamczyk}, \citenamefont {Adams}, \citenamefont {Adkins}, \citenamefont {Agakishiev}, \citenamefont {Aggarwal}, \citenamefont {Aggarwal} \emph {et~al.}}]{PhysRevC.104.L061901}%
  \BibitemOpen
  \bibfield  {author} {\bibinfo {author} {\bibfnamefont {M.~S.}\ \bibnamefont {Abdallah}}, \bibinfo {author} {\bibfnamefont {B.~E.}\ \bibnamefont {Aboona}}, \bibinfo {author} {\bibfnamefont {J.}~\bibnamefont {Adam}}, \bibinfo {author} {\bibfnamefont {L.}~\bibnamefont {Adamczyk}}, \bibinfo {author} {\bibfnamefont {J.~R.}\ \bibnamefont {Adams}}, \bibinfo {author} {\bibfnamefont {J.~K.}\ \bibnamefont {Adkins}}, \bibinfo {author} {\bibfnamefont {G.}~\bibnamefont {Agakishiev}}, \bibinfo {author} {\bibfnamefont {I.}~\bibnamefont {Aggarwal}}, \bibinfo {author} {\bibfnamefont {M.~M.}\ \bibnamefont {Aggarwal}}, \emph {et~al.} (\bibinfo {collaboration} {STAR Collaboration}),\ }\bibfield  {title} {\bibinfo {title} {Global $\mathrm{\ensuremath{\Lambda}}$-hyperon polarization in $\mathrm{Au}+\mathrm{Au}$ collisions at $\sqrt{{s}_{\mathrm{NN}}}=3 \mathrm{GeV}$},\ }\href {https://doi.org/10.1103/PhysRevC.104.L061901} {\bibfield  {journal} {\bibinfo  {journal} {Phys. Rev. C}\ }\textbf {\bibinfo {volume} {104}},\ \bibinfo
  {pages} {L061901} (\bibinfo {year} {2021})}\BibitemShut {NoStop}%
\bibitem [{\citenamefont {Abdulhamid}\ \emph {et~al.}(2023)\citenamefont {Abdulhamid}, \citenamefont {Aboona}, \citenamefont {Adam}, \citenamefont {Adamczyk}, \citenamefont {Adams}, \citenamefont {Aggarwal}, \citenamefont {Aggarwal}, \citenamefont {Ahammed}, \citenamefont {Alpatov} \emph {et~al.}}]{PhysRevC.108.014910}%
  \BibitemOpen
  \bibfield  {author} {\bibinfo {author} {\bibfnamefont {M.~I.}\ \bibnamefont {Abdulhamid}}, \bibinfo {author} {\bibfnamefont {B.~E.}\ \bibnamefont {Aboona}}, \bibinfo {author} {\bibfnamefont {J.}~\bibnamefont {Adam}}, \bibinfo {author} {\bibfnamefont {L.}~\bibnamefont {Adamczyk}}, \bibinfo {author} {\bibfnamefont {J.~R.}\ \bibnamefont {Adams}}, \bibinfo {author} {\bibfnamefont {I.}~\bibnamefont {Aggarwal}}, \bibinfo {author} {\bibfnamefont {M.~M.}\ \bibnamefont {Aggarwal}}, \bibinfo {author} {\bibfnamefont {Z.}~\bibnamefont {Ahammed}}, \bibinfo {author} {\bibfnamefont {E.}~\bibnamefont {Alpatov}}, \emph {et~al.} (\bibinfo {collaboration} {STAR Collaboration}),\ }\bibfield  {title} {\bibinfo {title} {Global polarization of $\mathrm{\ensuremath{\Lambda}}$ and $\overline{\mathrm{\ensuremath{\Lambda}}}$ hyperons in $\mathrm{Au}+\mathrm{Au}$ collisions at $\sqrt{{s}_{NN}}=19.6$ and 27 gev},\ }\href {https://doi.org/10.1103/PhysRevC.108.014910} {\bibfield  {journal} {\bibinfo  {journal} {Phys. Rev. C}\ }\textbf
  {\bibinfo {volume} {108}},\ \bibinfo {pages} {014910} (\bibinfo {year} {2023})}\BibitemShut {NoStop}%
\bibitem [{\citenamefont {Montenegro}\ \emph {et~al.}(2017)\citenamefont {Montenegro}, \citenamefont {Tinti},\ and\ \citenamefont {Torrieri}}]{Montenegro_2017}%
  \BibitemOpen
  \bibfield  {author} {\bibinfo {author} {\bibfnamefont {D.}~\bibnamefont {Montenegro}}, \bibinfo {author} {\bibfnamefont {L.}~\bibnamefont {Tinti}},\ and\ \bibinfo {author} {\bibfnamefont {G.}~\bibnamefont {Torrieri}},\ }\bibfield  {title} {\bibinfo {title} {Ideal relativistic fluid limit for a medium with polarization},\ }\bibfield  {journal} {\bibinfo  {journal} {Physical Review D}\ }\textbf {\bibinfo {volume} {96}},\ \href {https://doi.org/10.1103/physrevd.96.056012} {10.1103/physrevd.96.056012} (\bibinfo {year} {2017})\BibitemShut {NoStop}%
\bibitem [{\citenamefont {Florkowski}\ \emph {et~al.}(2018)\citenamefont {Florkowski}, \citenamefont {Friman}, \citenamefont {Jaiswal},\ and\ \citenamefont {Speranza}}]{PhysRevC.97.041901}%
  \BibitemOpen
  \bibfield  {author} {\bibinfo {author} {\bibfnamefont {W.}~\bibnamefont {Florkowski}}, \bibinfo {author} {\bibfnamefont {B.}~\bibnamefont {Friman}}, \bibinfo {author} {\bibfnamefont {A.}~\bibnamefont {Jaiswal}},\ and\ \bibinfo {author} {\bibfnamefont {E.}~\bibnamefont {Speranza}},\ }\bibfield  {title} {\bibinfo {title} {Relativistic fluid dynamics with spin},\ }\href {https://doi.org/10.1103/PhysRevC.97.041901} {\bibfield  {journal} {\bibinfo  {journal} {Phys. Rev. C}\ }\textbf {\bibinfo {volume} {97}},\ \bibinfo {pages} {041901} (\bibinfo {year} {2018})}\BibitemShut {NoStop}%
\bibitem [{\citenamefont {Becattini}\ \emph {et~al.}(2019)\citenamefont {Becattini}, \citenamefont {Florkowski},\ and\ \citenamefont {Speranza}}]{BECATTINI2019419}%
  \BibitemOpen
  \bibfield  {author} {\bibinfo {author} {\bibfnamefont {F.}~\bibnamefont {Becattini}}, \bibinfo {author} {\bibfnamefont {W.}~\bibnamefont {Florkowski}},\ and\ \bibinfo {author} {\bibfnamefont {E.}~\bibnamefont {Speranza}},\ }\bibfield  {title} {\bibinfo {title} {Spin tensor and its role in non-equilibrium thermodynamics},\ }\href {https://doi.org/https://doi.org/10.1016/j.physletb.2018.12.016} {\bibfield  {journal} {\bibinfo  {journal} {Physics Letters B}\ }\textbf {\bibinfo {volume} {789}},\ \bibinfo {pages} {419} (\bibinfo {year} {2019})}\BibitemShut {NoStop}%
\bibitem [{\citenamefont {Florkowski}(2019)}]{Florkowski_2019}%
  \BibitemOpen
  \bibfield  {author} {\bibinfo {author} {\bibfnamefont {W.}~\bibnamefont {Florkowski}},\ }\bibfield  {title} {\bibinfo {title} {Hydrodynamics with spin --- recent developments},\ }\href {https://doi.org/10.5506/aphyspolb.50.1047} {\bibfield  {journal} {\bibinfo  {journal} {Acta Physica Polonica B}\ }\textbf {\bibinfo {volume} {50}},\ \bibinfo {pages} {1047} (\bibinfo {year} {2019})}\BibitemShut {NoStop}%
\bibitem [{\citenamefont {Montenegro}\ and\ \citenamefont {Torrieri}(2020)}]{PhysRevD.102.036007}%
  \BibitemOpen
  \bibfield  {author} {\bibinfo {author} {\bibfnamefont {D.}~\bibnamefont {Montenegro}}\ and\ \bibinfo {author} {\bibfnamefont {G.}~\bibnamefont {Torrieri}},\ }\bibfield  {title} {\bibinfo {title} {Linear response theory and effective action of relativistic hydrodynamics with spin},\ }\href {https://doi.org/10.1103/PhysRevD.102.036007} {\bibfield  {journal} {\bibinfo  {journal} {Phys. Rev. D}\ }\textbf {\bibinfo {volume} {102}},\ \bibinfo {pages} {036007} (\bibinfo {year} {2020})}\BibitemShut {NoStop}%
\bibitem [{\citenamefont {Gallegos}\ and\ \citenamefont {G{\"u}rsoy}(2020)}]{Gallegos2020}%
  \BibitemOpen
  \bibfield  {author} {\bibinfo {author} {\bibfnamefont {A.~D.}\ \bibnamefont {Gallegos}}\ and\ \bibinfo {author} {\bibfnamefont {U.}~\bibnamefont {G{\"u}rsoy}},\ }\bibfield  {title} {\bibinfo {title} {Holographic spin liquids and lovelock chern-simons gravity},\ }\href {https://doi.org/10.1007/JHEP11(2020)151} {\bibfield  {journal} {\bibinfo  {journal} {Journal of High Energy Physics}\ }\textbf {\bibinfo {volume} {2020}},\ \bibinfo {pages} {151} (\bibinfo {year} {2020})}\BibitemShut {NoStop}%
\bibitem [{\citenamefont {Garbiso}\ and\ \citenamefont {Kaminski}(2020)}]{Garbiso2020}%
  \BibitemOpen
  \bibfield  {author} {\bibinfo {author} {\bibfnamefont {M.}~\bibnamefont {Garbiso}}\ and\ \bibinfo {author} {\bibfnamefont {M.}~\bibnamefont {Kaminski}},\ }\bibfield  {title} {\bibinfo {title} {Hydrodynamics of simply spinning black holes {\&} hydrodynamics for spinning quantum fluids},\ }\href {https://doi.org/10.1007/JHEP12(2020)112} {\bibfield  {journal} {\bibinfo  {journal} {Journal of High Energy Physics}\ }\textbf {\bibinfo {volume} {2020}},\ \bibinfo {pages} {112} (\bibinfo {year} {2020})}\BibitemShut {NoStop}%
\bibitem [{\citenamefont {Gallegos}\ \emph {et~al.}(2021)\citenamefont {Gallegos}, \citenamefont {Gursoy},\ and\ \citenamefont {Yarom}}]{10.21468/SciPostPhys.11.2.041}%
  \BibitemOpen
  \bibfield  {author} {\bibinfo {author} {\bibfnamefont {D.}~\bibnamefont {Gallegos}}, \bibinfo {author} {\bibfnamefont {U.}~\bibnamefont {Gursoy}},\ and\ \bibinfo {author} {\bibfnamefont {A.}~\bibnamefont {Yarom}},\ }\bibfield  {title} {\bibinfo {title} {{Hydrodynamics of spin currents}},\ }\href {https://doi.org/10.21468/SciPostPhys.11.2.041} {\bibfield  {journal} {\bibinfo  {journal} {SciPost Phys.}\ }\textbf {\bibinfo {volume} {11}},\ \bibinfo {pages} {041} (\bibinfo {year} {2021})}\BibitemShut {NoStop}%
\bibitem [{\citenamefont {Becattini}(2021)}]{Becattini_2021}%
  \BibitemOpen
  \bibfield  {author} {\bibinfo {author} {\bibfnamefont {F.}~\bibnamefont {Becattini}},\ }\bibfield  {title} {\bibinfo {title} {Does the spin tensor play any role in non-gravitational physics?},\ }\href {https://doi.org/10.1016/j.nuclphysa.2020.121833} {\bibfield  {journal} {\bibinfo  {journal} {Nuclear Physics A}\ }\textbf {\bibinfo {volume} {1005}},\ \bibinfo {pages} {121833} (\bibinfo {year} {2021})}\BibitemShut {NoStop}%
\bibitem [{\citenamefont {Hongo}\ \emph {et~al.}(2021)\citenamefont {Hongo}, \citenamefont {Huang}, \citenamefont {Kaminski}, \citenamefont {Stephanov},\ and\ \citenamefont {Yee}}]{Hongo2021}%
  \BibitemOpen
  \bibfield  {author} {\bibinfo {author} {\bibfnamefont {M.}~\bibnamefont {Hongo}}, \bibinfo {author} {\bibfnamefont {X.-G.}\ \bibnamefont {Huang}}, \bibinfo {author} {\bibfnamefont {M.}~\bibnamefont {Kaminski}}, \bibinfo {author} {\bibfnamefont {M.}~\bibnamefont {Stephanov}},\ and\ \bibinfo {author} {\bibfnamefont {H.-U.}\ \bibnamefont {Yee}},\ }\bibfield  {title} {\bibinfo {title} {Relativistic spin hydrodynamics with torsion and linear response theory for spin relaxation},\ }\href {https://doi.org/10.1007/JHEP11(2021)150} {\bibfield  {journal} {\bibinfo  {journal} {Journal of High Energy Physics}\ }\textbf {\bibinfo {volume} {2021}},\ \bibinfo {pages} {150} (\bibinfo {year} {2021})}\BibitemShut {NoStop}%
\bibitem [{\citenamefont {Das}\ \emph {et~al.}(2021)\citenamefont {Das}, \citenamefont {Florkowski}, \citenamefont {Ryblewski},\ and\ \citenamefont {Singh}}]{Das_2021}%
  \BibitemOpen
  \bibfield  {author} {\bibinfo {author} {\bibfnamefont {A.}~\bibnamefont {Das}}, \bibinfo {author} {\bibfnamefont {W.}~\bibnamefont {Florkowski}}, \bibinfo {author} {\bibfnamefont {R.}~\bibnamefont {Ryblewski}},\ and\ \bibinfo {author} {\bibfnamefont {R.}~\bibnamefont {Singh}},\ }\bibfield  {title} {\bibinfo {title} {Pseudogauge dependence of quantum fluctuations of the energy in a hot relativistic gas of fermions},\ }\bibfield  {journal} {\bibinfo  {journal} {Physical Review D}\ }\textbf {\bibinfo {volume} {103}},\ \href {https://doi.org/10.1103/physrevd.103.l091502} {10.1103/physrevd.103.l091502} (\bibinfo {year} {2021})\BibitemShut {NoStop}%
\bibitem [{\citenamefont {Floerchinger}\ and\ \citenamefont {Grossi}(2022)}]{Floerchinger_2022}%
  \BibitemOpen
  \bibfield  {author} {\bibinfo {author} {\bibfnamefont {S.}~\bibnamefont {Floerchinger}}\ and\ \bibinfo {author} {\bibfnamefont {E.}~\bibnamefont {Grossi}},\ }\bibfield  {title} {\bibinfo {title} {Conserved and nonconserved noether currents from the quantum effective action},\ }\bibfield  {journal} {\bibinfo  {journal} {Physical Review D}\ }\textbf {\bibinfo {volume} {105}},\ \href {https://doi.org/10.1103/physrevd.105.085015} {10.1103/physrevd.105.085015} (\bibinfo {year} {2022})\BibitemShut {NoStop}%
\bibitem [{\citenamefont {Weickgenannt}\ \emph {et~al.}(2022)\citenamefont {Weickgenannt}, \citenamefont {Wagner},\ and\ \citenamefont {Speranza}}]{Weickgenannt_2022}%
  \BibitemOpen
  \bibfield  {author} {\bibinfo {author} {\bibfnamefont {N.}~\bibnamefont {Weickgenannt}}, \bibinfo {author} {\bibfnamefont {D.}~\bibnamefont {Wagner}},\ and\ \bibinfo {author} {\bibfnamefont {E.}~\bibnamefont {Speranza}},\ }\bibfield  {title} {\bibinfo {title} {Pseudogauges and relativistic spin hydrodynamics for interacting dirac and proca fields},\ }\bibfield  {journal} {\bibinfo  {journal} {Physical Review D}\ }\textbf {\bibinfo {volume} {105}},\ \href {https://doi.org/10.1103/physrevd.105.116026} {10.1103/physrevd.105.116026} (\bibinfo {year} {2022})\BibitemShut {NoStop}%
\bibitem [{\citenamefont {Buzzegoli}(2022)}]{Buzzegoli_2022}%
  \BibitemOpen
  \bibfield  {author} {\bibinfo {author} {\bibfnamefont {M.}~\bibnamefont {Buzzegoli}},\ }\bibfield  {title} {\bibinfo {title} {Pseudogauge dependence of the spin polarization and of the axial vortical effect},\ }\bibfield  {journal} {\bibinfo  {journal} {Physical Review C}\ }\textbf {\bibinfo {volume} {105}},\ \href {https://doi.org/10.1103/physrevc.105.044907} {10.1103/physrevc.105.044907} (\bibinfo {year} {2022})\BibitemShut {NoStop}%
\bibitem [{\citenamefont {Gallegos}\ \emph {et~al.}(2023)\citenamefont {Gallegos}, \citenamefont {Gürsoy},\ and\ \citenamefont {Yarom}}]{Gallegos_2023}%
  \BibitemOpen
  \bibfield  {author} {\bibinfo {author} {\bibfnamefont {A.~D.}\ \bibnamefont {Gallegos}}, \bibinfo {author} {\bibfnamefont {U.}~\bibnamefont {Gürsoy}},\ and\ \bibinfo {author} {\bibfnamefont {A.}~\bibnamefont {Yarom}},\ }\bibfield  {title} {\bibinfo {title} {Hydrodynamics, spin currents and torsion},\ }\bibfield  {journal} {\bibinfo  {journal} {Journal of High Energy Physics}\ }\textbf {\bibinfo {volume} {2023}},\ \href {https://doi.org/10.1007/jhep05(2023)139} {10.1007/jhep05(2023)139} (\bibinfo {year} {2023})\BibitemShut {NoStop}%
\bibitem [{\citenamefont {Dey}\ \emph {et~al.}(2024)\citenamefont {Dey}, \citenamefont {Bhadury}, \citenamefont {Florkowski}, \citenamefont {Ryblewski},\ and\ \citenamefont {Jaiswal}}]{dey2024energymomentumcorrelatorsfermionsfinite}%
  \BibitemOpen
  \bibfield  {author} {\bibinfo {author} {\bibfnamefont {S.}~\bibnamefont {Dey}}, \bibinfo {author} {\bibfnamefont {S.}~\bibnamefont {Bhadury}}, \bibinfo {author} {\bibfnamefont {W.}~\bibnamefont {Florkowski}}, \bibinfo {author} {\bibfnamefont {R.}~\bibnamefont {Ryblewski}},\ and\ \bibinfo {author} {\bibfnamefont {A.}~\bibnamefont {Jaiswal}},\ }\href {https://arxiv.org/abs/2409.18912} {\bibinfo {title} {Energy-momentum correlators of fermions at finite temperature and density}} (\bibinfo {year} {2024}),\ \Eprint {https://arxiv.org/abs/2409.18912} {arXiv:2409.18912 [hep-ph]} \BibitemShut {NoStop}%
\bibitem [{\citenamefont {Drogosz}\ \emph {et~al.}(2025)\citenamefont {Drogosz}, \citenamefont {Florkowski}, \citenamefont {Hontarenko},\ and\ \citenamefont {Ryblewski}}]{Drogosz_2025}%
  \BibitemOpen
  \bibfield  {author} {\bibinfo {author} {\bibfnamefont {Z.}~\bibnamefont {Drogosz}}, \bibinfo {author} {\bibfnamefont {W.}~\bibnamefont {Florkowski}}, \bibinfo {author} {\bibfnamefont {M.}~\bibnamefont {Hontarenko}},\ and\ \bibinfo {author} {\bibfnamefont {R.}~\bibnamefont {Ryblewski}},\ }\bibfield  {title} {\bibinfo {title} {Dynamical constraints on pseudo-gauge transformations},\ }\href {https://doi.org/10.1016/j.physletb.2025.139244} {\bibfield  {journal} {\bibinfo  {journal} {Physics Letters B}\ }\textbf {\bibinfo {volume} {861}},\ \bibinfo {pages} {139244} (\bibinfo {year} {2025})}\BibitemShut {NoStop}%
\bibitem [{\citenamefont {Daher}\ \emph {et~al.}(2025)\citenamefont {Daher}, \citenamefont {Sheng}, \citenamefont {Wagner},\ and\ \citenamefont {Becattini}}]{Daher:2025pfq}%
  \BibitemOpen
  \bibfield  {author} {\bibinfo {author} {\bibfnamefont {A.}~\bibnamefont {Daher}}, \bibinfo {author} {\bibfnamefont {X.-L.}\ \bibnamefont {Sheng}}, \bibinfo {author} {\bibfnamefont {D.}~\bibnamefont {Wagner}},\ and\ \bibinfo {author} {\bibfnamefont {F.}~\bibnamefont {Becattini}},\ }\bibfield  {title} {\bibinfo {title} {{Dissipative currents and transport coefficients in relativistic spin hydrodynamics}},\ }\href {https://doi.org/10.1103/zttq-cs4l} {\bibfield  {journal} {\bibinfo  {journal} {Phys. Rev. D}\ }\textbf {\bibinfo {volume} {112}},\ \bibinfo {pages} {094020} (\bibinfo {year} {2025})},\ \Eprint {https://arxiv.org/abs/2503.03713} {arXiv:2503.03713 [nucl-th]} \BibitemShut {NoStop}%
\bibitem [{\citenamefont {Becattini}\ and\ \citenamefont {Singh}(2025)}]{Becattini:2025oyi}%
  \BibitemOpen
  \bibfield  {author} {\bibinfo {author} {\bibfnamefont {F.}~\bibnamefont {Becattini}}\ and\ \bibinfo {author} {\bibfnamefont {R.}~\bibnamefont {Singh}},\ }\bibfield  {title} {\bibinfo {title} {{On the local thermodynamic relations in relativistic spin hydrodynamics}},\ }\href {https://doi.org/10.1140/epjc/s10052-025-15071-3} {\bibfield  {journal} {\bibinfo  {journal} {Eur. Phys. J. C}\ }\textbf {\bibinfo {volume} {85}},\ \bibinfo {pages} {1338} (\bibinfo {year} {2025})},\ \Eprint {https://arxiv.org/abs/2506.20681} {arXiv:2506.20681 [nucl-th]} \BibitemShut {NoStop}%
\bibitem [{\citenamefont {Zhang}\ \emph {et~al.}(2025)\citenamefont {Zhang}, \citenamefont {Huang}, \citenamefont {Becattini},\ and\ \citenamefont {Sheng}}]{Zhang:2024mhs}%
  \BibitemOpen
  \bibfield  {author} {\bibinfo {author} {\bibfnamefont {Z.-H.}\ \bibnamefont {Zhang}}, \bibinfo {author} {\bibfnamefont {X.-G.}\ \bibnamefont {Huang}}, \bibinfo {author} {\bibfnamefont {F.}~\bibnamefont {Becattini}},\ and\ \bibinfo {author} {\bibfnamefont {X.-L.}\ \bibnamefont {Sheng}},\ }\bibfield  {title} {\bibinfo {title} {{Vector and tensor spin polarization for vector bosons at local equilibrium}},\ }\href {https://doi.org/10.1007/JHEP07(2025)224} {\bibfield  {journal} {\bibinfo  {journal} {JHEP}\ }\textbf {\bibinfo {volume} {07}},\ \bibinfo {pages} {224}},\ \Eprint {https://arxiv.org/abs/2412.19416} {arXiv:2412.19416 [hep-ph]} \BibitemShut {NoStop}%
\bibitem [{\citenamefont {Cartwright}\ \emph {et~al.}(2025)\citenamefont {Cartwright}, \citenamefont {Gallegos}, \citenamefont {G{\"u}rsoy}, \citenamefont {Klein},\ and\ \citenamefont {Yarom}}]{Cartwright2025}%
  \BibitemOpen
  \bibfield  {author} {\bibinfo {author} {\bibfnamefont {C.}~\bibnamefont {Cartwright}}, \bibinfo {author} {\bibfnamefont {D.}~\bibnamefont {Gallegos}}, \bibinfo {author} {\bibfnamefont {U.}~\bibnamefont {G{\"u}rsoy}}, \bibinfo {author} {\bibfnamefont {R.}~\bibnamefont {Klein}},\ and\ \bibinfo {author} {\bibfnamefont {A.}~\bibnamefont {Yarom}},\ }\bibfield  {title} {\bibinfo {title} {A supersymmetric spin current},\ }\href {https://doi.org/10.1007/JHEP08(2025)129} {\bibfield  {journal} {\bibinfo  {journal} {Journal of High Energy Physics}\ }\textbf {\bibinfo {volume} {2025}},\ \bibinfo {pages} {129} (\bibinfo {year} {2025})}\BibitemShut {NoStop}%
\bibitem [{\citenamefont {Belinfante}(1940)}]{BELINFANTE1940449}%
  \BibitemOpen
  \bibfield  {author} {\bibinfo {author} {\bibfnamefont {F.}~\bibnamefont {Belinfante}},\ }\bibfield  {title} {\bibinfo {title} {On the current and the density of the electric charge, the energy, the linear momentum and the angular momentum of arbitrary fields},\ }\href {https://doi.org/https://doi.org/10.1016/S0031-8914(40)90091-X} {\bibfield  {journal} {\bibinfo  {journal} {Physica}\ }\textbf {\bibinfo {volume} {7}},\ \bibinfo {pages} {449} (\bibinfo {year} {1940})}\BibitemShut {NoStop}%
\bibitem [{\citenamefont {Rosenfeld}(1940)}]{Rosenfeld1940}%
  \BibitemOpen
  \bibfield  {author} {\bibinfo {author} {\bibfnamefont {L.}~\bibnamefont {Rosenfeld}},\ }\bibfield  {title} {\bibinfo {title} {Sur le tenseur d'impulsion-{\'e}nergie},\ }\href@noop {} {\bibfield  {journal} {\bibinfo  {journal} {M{\'e}moires Acad. Roy. De Belgique. 18 (6):1–30.}\ } (\bibinfo {year} {1940})}\BibitemShut {NoStop}%
\bibitem [{\citenamefont {Blaschke}\ \emph {et~al.}(2016)\citenamefont {Blaschke}, \citenamefont {Gieres}, \citenamefont {Reboud},\ and\ \citenamefont {Schweda}}]{Blaschke_2016}%
  \BibitemOpen
  \bibfield  {author} {\bibinfo {author} {\bibfnamefont {D.~N.}\ \bibnamefont {Blaschke}}, \bibinfo {author} {\bibfnamefont {F.}~\bibnamefont {Gieres}}, \bibinfo {author} {\bibfnamefont {M.}~\bibnamefont {Reboud}},\ and\ \bibinfo {author} {\bibfnamefont {M.}~\bibnamefont {Schweda}},\ }\bibfield  {title} {\bibinfo {title} {The energy–momentum tensor(s) in classical gauge theories},\ }\href {https://doi.org/10.1016/j.nuclphysb.2016.07.001} {\bibfield  {journal} {\bibinfo  {journal} {Nuclear Physics B}\ }\textbf {\bibinfo {volume} {912}},\ \bibinfo {pages} {192–223} (\bibinfo {year} {2016})}\BibitemShut {NoStop}%
\bibitem [{\citenamefont {Speranza}\ and\ \citenamefont {Weickgenannt}(2021)}]{Speranza_2021}%
  \BibitemOpen
  \bibfield  {author} {\bibinfo {author} {\bibfnamefont {E.}~\bibnamefont {Speranza}}\ and\ \bibinfo {author} {\bibfnamefont {N.}~\bibnamefont {Weickgenannt}},\ }\bibfield  {title} {\bibinfo {title} {Spin tensor and pseudo-gauges: from nuclear collisions to gravitational physics},\ }\bibfield  {journal} {\bibinfo  {journal} {The European Physical Journal A}\ }\textbf {\bibinfo {volume} {57}},\ \href {https://doi.org/10.1140/epja/s10050-021-00455-2} {10.1140/epja/s10050-021-00455-2} (\bibinfo {year} {2021})\BibitemShut {NoStop}%
\bibitem [{\citenamefont {Becattini}\ and\ \citenamefont {Hoyos}(2025)}]{becattini2025pseudogauge}%
  \BibitemOpen
  \bibfield  {author} {\bibinfo {author} {\bibfnamefont {F.}~\bibnamefont {Becattini}}\ and\ \bibinfo {author} {\bibfnamefont {C.}~\bibnamefont {Hoyos}},\ }\href {https://arxiv.org/abs/2507.09249} {\bibinfo {title} {Pseudo-gauge invariant non-equilibrium density operator}} (\bibinfo {year} {2025}),\ \Eprint {https://arxiv.org/abs/2507.09249} {arXiv:2507.09249 [nucl-th]} \BibitemShut {NoStop}%
\bibitem [{\citenamefont {Armas}\ and\ \citenamefont {Jain}(2026)}]{armas2026}%
  \BibitemOpen
  \bibfield  {author} {\bibinfo {author} {\bibfnamefont {J.}~\bibnamefont {Armas}}\ and\ \bibinfo {author} {\bibfnamefont {A.}~\bibnamefont {Jain}},\ }\href {https://arxiv.org/abs/2601.14421} {\bibinfo {title} {Thermodynamics of ideal spin fluids and pseudo-gauge ambiguity}} (\bibinfo {year} {2026}),\ \Eprint {https://arxiv.org/abs/2601.14421} {arXiv:2601.14421 [hep-th]} \BibitemShut {NoStop}%
\bibitem [{\citenamefont {Jeon}(2023)}]{Jeon_2023}%
  \BibitemOpen
  \bibfield  {author} {\bibinfo {author} {\bibfnamefont {S.}~\bibnamefont {Jeon}},\ }\bibfield  {title} {\bibinfo {title} {Conservation laws and hydrodynamics},\ }\href {https://doi.org/10.1051/epjconf/202327601010} {\bibfield  {journal} {\bibinfo  {journal} {EPJ Web of Conferences}\ }\textbf {\bibinfo {volume} {276}},\ \bibinfo {pages} {01010} (\bibinfo {year} {2023})}\BibitemShut {NoStop}%
\bibitem [{\citenamefont {Buzzegoli}\ and\ \citenamefont {Palermo}(2024)}]{PhysRevLett.133.262301}%
  \BibitemOpen
  \bibfield  {author} {\bibinfo {author} {\bibfnamefont {M.}~\bibnamefont {Buzzegoli}}\ and\ \bibinfo {author} {\bibfnamefont {A.}~\bibnamefont {Palermo}},\ }\bibfield  {title} {\bibinfo {title} {Emergent canonical spin tensor in the chiral-symmetric hot qcd},\ }\href {https://doi.org/10.1103/PhysRevLett.133.262301} {\bibfield  {journal} {\bibinfo  {journal} {Phys. Rev. Lett.}\ }\textbf {\bibinfo {volume} {133}},\ \bibinfo {pages} {262301} (\bibinfo {year} {2024})}\BibitemShut {NoStop}%
\bibitem [{\citenamefont {Iosifidis}\ \emph {et~al.}(2025)\citenamefont {Iosifidis}, \citenamefont {Karydas}, \citenamefont {Petkou},\ and\ \citenamefont {Siampos}}]{Iosifidis_2025}%
  \BibitemOpen
  \bibfield  {author} {\bibinfo {author} {\bibfnamefont {D.}~\bibnamefont {Iosifidis}}, \bibinfo {author} {\bibfnamefont {M.}~\bibnamefont {Karydas}}, \bibinfo {author} {\bibfnamefont {A.}~\bibnamefont {Petkou}},\ and\ \bibinfo {author} {\bibfnamefont {K.}~\bibnamefont {Siampos}},\ }\bibfield  {title} {\bibinfo {title} {Geometric origin of the energy-momentum tensor improvement terms},\ }\bibfield  {journal} {\bibinfo  {journal} {Physical Review D}\ }\textbf {\bibinfo {volume} {112}},\ \href {https://doi.org/10.1103/wby2-d33f} {10.1103/wby2-d33f} (\bibinfo {year} {2025})\BibitemShut {NoStop}%
\bibitem [{\citenamefont {Labastida}\ and\ \citenamefont {Morris}(1986)}]{Labastida:1986gy}%
  \BibitemOpen
  \bibfield  {author} {\bibinfo {author} {\bibfnamefont {J.~M.~F.}\ \bibnamefont {Labastida}}\ and\ \bibinfo {author} {\bibfnamefont {T.~R.}\ \bibnamefont {Morris}},\ }\bibfield  {title} {\bibinfo {title} {{MASSLESS MIXED SYMMETRY BOSONIC FREE FIELDS}},\ }\href {https://doi.org/10.1016/0370-2693(86)90143-7} {\bibfield  {journal} {\bibinfo  {journal} {Phys. Lett. B}\ }\textbf {\bibinfo {volume} {180}},\ \bibinfo {pages} {101} (\bibinfo {year} {1986})}\BibitemShut {NoStop}%
\bibitem [{\citenamefont {Dubois-Violette}(2000)}]{Dubois-Violette:2000fok}%
  \BibitemOpen
  \bibfield  {author} {\bibinfo {author} {\bibfnamefont {M.}~\bibnamefont {Dubois-Violette}},\ }\bibfield  {title} {\bibinfo {title} {{Lectures on differentials, generalized differentials and on some examples related to theoretical physics}},\ }in\ \href@noop {} {\emph {\bibinfo {booktitle} {{School: Bariloche 2000: Quantum Symmetries in Theoretical Physics and Mathematics}}}}\ (\bibinfo {year} {2000})\ \Eprint {https://arxiv.org/abs/math/0005256} {arXiv:math/0005256} \BibitemShut {NoStop}%
\bibitem [{\citenamefont {Dubois-Violette}\ and\ \citenamefont {Henneaux}(2002)}]{Dubois-Violette:2001wjr}%
  \BibitemOpen
  \bibfield  {author} {\bibinfo {author} {\bibfnamefont {M.}~\bibnamefont {Dubois-Violette}}\ and\ \bibinfo {author} {\bibfnamefont {M.}~\bibnamefont {Henneaux}},\ }\bibfield  {title} {\bibinfo {title} {{Tensor fields of mixed Young symmetry type and N complexes}},\ }\href {https://doi.org/10.1007/s002200200610} {\bibfield  {journal} {\bibinfo  {journal} {Commun. Math. Phys.}\ }\textbf {\bibinfo {volume} {226}},\ \bibinfo {pages} {393} (\bibinfo {year} {2002})},\ \Eprint {https://arxiv.org/abs/math/0110088} {arXiv:math/0110088} \BibitemShut {NoStop}%
\bibitem [{\citenamefont {de~Medeiros}\ and\ \citenamefont {Hull}(2003)}]{deMedeiros:2002qpr}%
  \BibitemOpen
  \bibfield  {author} {\bibinfo {author} {\bibfnamefont {P.}~\bibnamefont {de~Medeiros}}\ and\ \bibinfo {author} {\bibfnamefont {C.}~\bibnamefont {Hull}},\ }\bibfield  {title} {\bibinfo {title} {{Exotic tensor gauge theory and duality}},\ }\href {https://doi.org/10.1007/s00220-003-0810-z} {\bibfield  {journal} {\bibinfo  {journal} {Commun. Math. Phys.}\ }\textbf {\bibinfo {volume} {235}},\ \bibinfo {pages} {255} (\bibinfo {year} {2003})},\ \Eprint {https://arxiv.org/abs/hep-th/0208155} {arXiv:hep-th/0208155} \BibitemShut {NoStop}%
\bibitem [{\citenamefont {Bekaert}\ and\ \citenamefont {Boulanger}(2004)}]{Bekaert:2002dt}%
  \BibitemOpen
  \bibfield  {author} {\bibinfo {author} {\bibfnamefont {X.}~\bibnamefont {Bekaert}}\ and\ \bibinfo {author} {\bibfnamefont {N.}~\bibnamefont {Boulanger}},\ }\bibfield  {title} {\bibinfo {title} {{Tensor gauge fields in arbitrary representations of GL(D,R): Duality and Poincare lemma}},\ }\href {https://doi.org/10.1007/s00220-003-0995-1} {\bibfield  {journal} {\bibinfo  {journal} {Commun. Math. Phys.}\ }\textbf {\bibinfo {volume} {245}},\ \bibinfo {pages} {27} (\bibinfo {year} {2004})},\ \Eprint {https://arxiv.org/abs/hep-th/0208058} {arXiv:hep-th/0208058} \BibitemShut {NoStop}%
\bibitem [{\citenamefont {Bruce}\ and\ \citenamefont {Ibarguengoytia}(2019)}]{Bruce_2019}%
  \BibitemOpen
  \bibfield  {author} {\bibinfo {author} {\bibfnamefont {A.~J.}\ \bibnamefont {Bruce}}\ and\ \bibinfo {author} {\bibfnamefont {E.}~\bibnamefont {Ibarguengoytia}},\ }\bibfield  {title} {\bibinfo {title} {The graded differential geometry of mixed symmetry tensors},\ }\href {https://doi.org/10.5817/am2019-2-123} {\bibfield  {journal} {\bibinfo  {journal} {Archivum Mathematicum}\ ,\ \bibinfo {pages} {123–137}} (\bibinfo {year} {2019})}\BibitemShut {NoStop}%
\bibitem [{\citenamefont {Chatzistavrakidis}\ \emph {et~al.}(2020)\citenamefont {Chatzistavrakidis}, \citenamefont {Karagiannis},\ and\ \citenamefont {Schupp}}]{Chatzistavrakidis_2020}%
  \BibitemOpen
  \bibfield  {author} {\bibinfo {author} {\bibfnamefont {A.}~\bibnamefont {Chatzistavrakidis}}, \bibinfo {author} {\bibfnamefont {G.}~\bibnamefont {Karagiannis}},\ and\ \bibinfo {author} {\bibfnamefont {P.}~\bibnamefont {Schupp}},\ }\bibfield  {title} {\bibinfo {title} {A unified approach to standard and exotic dualizations through graded geometry},\ }\href {https://doi.org/10.1007/s00220-020-03728-x} {\bibfield  {journal} {\bibinfo  {journal} {Communications in Mathematical Physics}\ }\textbf {\bibinfo {volume} {378}},\ \bibinfo {pages} {1157–1201} (\bibinfo {year} {2020})}\BibitemShut {NoStop}%
\bibitem [{\citenamefont {Hinterbichler}\ \emph {et~al.}(2023)\citenamefont {Hinterbichler}, \citenamefont {Hofman}, \citenamefont {Joyce},\ and\ \citenamefont {Mathys}}]{Hinterbichler:2022agn}%
  \BibitemOpen
  \bibfield  {author} {\bibinfo {author} {\bibfnamefont {K.}~\bibnamefont {Hinterbichler}}, \bibinfo {author} {\bibfnamefont {D.~M.}\ \bibnamefont {Hofman}}, \bibinfo {author} {\bibfnamefont {A.}~\bibnamefont {Joyce}},\ and\ \bibinfo {author} {\bibfnamefont {G.}~\bibnamefont {Mathys}},\ }\bibfield  {title} {\bibinfo {title} {{Gravity as a gapless phase and biform symmetries}},\ }\href {https://doi.org/10.1007/JHEP02(2023)151} {\bibfield  {journal} {\bibinfo  {journal} {JHEP}\ }\textbf {\bibinfo {volume} {02}},\ \bibinfo {pages} {151}},\ \Eprint {https://arxiv.org/abs/2205.12272} {arXiv:2205.12272 [hep-th]} \BibitemShut {NoStop}%
\bibitem [{\citenamefont {Hull}\ \emph {et~al.}(2025)\citenamefont {Hull}, \citenamefont {Hutt},\ and\ \citenamefont {Lindstr{\"o}m}}]{Hull:2024bcl}%
  \BibitemOpen
  \bibfield  {author} {\bibinfo {author} {\bibfnamefont {C.}~\bibnamefont {Hull}}, \bibinfo {author} {\bibfnamefont {M.~L.}\ \bibnamefont {Hutt}},\ and\ \bibinfo {author} {\bibfnamefont {U.}~\bibnamefont {Lindstr{\"o}m}},\ }\bibfield  {title} {\bibinfo {title} {{Generalised symmetries in linear gravity}},\ }\href {https://doi.org/10.1007/JHEP04(2025)046} {\bibfield  {journal} {\bibinfo  {journal} {JHEP}\ }\textbf {\bibinfo {volume} {04}},\ \bibinfo {pages} {046}},\ \Eprint {https://arxiv.org/abs/2409.00178} {arXiv:2409.00178 [hep-th]} \BibitemShut {NoStop}%
\bibitem [{\citenamefont {Freedman}\ and\ \citenamefont {Van~Proeyen}(2012)}]{Freedman:2012zz}%
  \BibitemOpen
  \bibfield  {author} {\bibinfo {author} {\bibfnamefont {D.~Z.}\ \bibnamefont {Freedman}}\ and\ \bibinfo {author} {\bibfnamefont {A.}~\bibnamefont {Van~Proeyen}},\ }\href {https://doi.org/10.1017/CBO9781139026833} {\emph {\bibinfo {title} {{Supergravity}}}}\ (\bibinfo  {publisher} {Cambridge Univ. Press},\ \bibinfo {address} {Cambridge, UK},\ \bibinfo {year} {2012})\BibitemShut {NoStop}%
\bibitem [{\citenamefont {Hehl}\ \emph {et~al.}(1995)\citenamefont {Hehl}, \citenamefont {McCrea}, \citenamefont {Mielke},\ and\ \citenamefont {Ne'eman}}]{Hehl:1994ue}%
  \BibitemOpen
  \bibfield  {author} {\bibinfo {author} {\bibfnamefont {F.~W.}\ \bibnamefont {Hehl}}, \bibinfo {author} {\bibfnamefont {J.~D.}\ \bibnamefont {McCrea}}, \bibinfo {author} {\bibfnamefont {E.~W.}\ \bibnamefont {Mielke}},\ and\ \bibinfo {author} {\bibfnamefont {Y.}~\bibnamefont {Ne'eman}},\ }\bibfield  {title} {\bibinfo {title} {{Metric affine gauge theory of gravity: Field equations, Noether identities, world spinors, and breaking of dilation invariance}},\ }\href {https://doi.org/10.1016/0370-1573(94)00111-F} {\bibfield  {journal} {\bibinfo  {journal} {Phys. Rept.}\ }\textbf {\bibinfo {volume} {258}},\ \bibinfo {pages} {1} (\bibinfo {year} {1995})},\ \Eprint {https://arxiv.org/abs/gr-qc/9402012} {arXiv:gr-qc/9402012} \BibitemShut {NoStop}%
\bibitem [{\citenamefont {Jiménez-Cano}(2022)}]{jiménezcano2022metricaffinegaugetheoriesgravity}%
  \BibitemOpen
  \bibfield  {author} {\bibinfo {author} {\bibfnamefont {A.}~\bibnamefont {Jiménez-Cano}},\ }\href {https://arxiv.org/abs/2201.12847} {\bibinfo {title} {Metric-affine gauge theories of gravity: Foundations and new insights}} (\bibinfo {year} {2022}),\ \Eprint {https://arxiv.org/abs/2201.12847} {arXiv:2201.12847 [gr-qc]} \BibitemShut {NoStop}%
\bibitem [{\citenamefont {Carroll}(2019)}]{Carroll_2019}%
  \BibitemOpen
  \bibfield  {author} {\bibinfo {author} {\bibfnamefont {S.~M.}\ \bibnamefont {Carroll}},\ }\href@noop {} {\emph {\bibinfo {title} {Spacetime and Geometry: An Introduction to General Relativity}}}\ (\bibinfo  {publisher} {Cambridge University Press},\ \bibinfo {year} {2019})\BibitemShut {NoStop}%
\bibitem [{\citenamefont {Hehl}(1976)}]{HEHL197655}%
  \BibitemOpen
  \bibfield  {author} {\bibinfo {author} {\bibfnamefont {F.~W.}\ \bibnamefont {Hehl}},\ }\bibfield  {title} {\bibinfo {title} {On the energy tensor of spinning massive matter in classical field theory and general relativity},\ }\href {https://doi.org/https://doi.org/10.1016/0034-4877(76)90016-1} {\bibfield  {journal} {\bibinfo  {journal} {Reports on Mathematical Physics}\ }\textbf {\bibinfo {volume} {9}},\ \bibinfo {pages} {55} (\bibinfo {year} {1976})}\BibitemShut {NoStop}%
\bibitem [{\citenamefont {Trautman}(2006)}]{Trautman:2006fp}%
  \BibitemOpen
  \bibfield  {author} {\bibinfo {author} {\bibfnamefont {A.}~\bibnamefont {Trautman}},\ }\href@noop {} {\bibinfo {title} {{Einstein-Cartan theory}}} (\bibinfo {year} {2006}),\ \Eprint {https://arxiv.org/abs/gr-qc/0606062} {arXiv:gr-qc/0606062} \BibitemShut {NoStop}%
\bibitem [{\citenamefont {Bertlmann}(1996)}]{Bertlmann:1996xk}%
  \BibitemOpen
  \bibfield  {author} {\bibinfo {author} {\bibfnamefont {R.~A.}\ \bibnamefont {Bertlmann}},\ }\href@noop {} {\emph {\bibinfo {title} {{Anomalies in quantum field theory}}}}\ (\bibinfo  {publisher} {Oxford University Press},\ \bibinfo {year} {1996})\BibitemShut {NoStop}%
\bibitem [{\citenamefont {Landau}\ and\ \citenamefont {Lifshitz}(1984)}]{landau}%
  \BibitemOpen
  \bibfield  {author} {\bibinfo {author} {\bibfnamefont {L.}~\bibnamefont {Landau}}\ and\ \bibinfo {author} {\bibfnamefont {E.}~\bibnamefont {Lifshitz}},\ }\href@noop {} {\emph {\bibinfo {title} {Electrodynamics of continuous media}}}\ (\bibinfo  {publisher} {Pergamon Press},\ \bibinfo {year} {1984})\BibitemShut {NoStop}%
\bibitem [{\citenamefont {Jackson}(1998)}]{Jackson:1998nia}%
  \BibitemOpen
  \bibfield  {author} {\bibinfo {author} {\bibfnamefont {J.~D.}\ \bibnamefont {Jackson}},\ }\href@noop {} {\emph {\bibinfo {title} {{Classical Electrodynamics}}}}\ (\bibinfo  {publisher} {Wiley},\ \bibinfo {year} {1998})\BibitemShut {NoStop}%
\bibitem [{\citenamefont {Kovtun}(2016)}]{Kovtun_2016}%
  \BibitemOpen
  \bibfield  {author} {\bibinfo {author} {\bibfnamefont {P.}~\bibnamefont {Kovtun}},\ }\bibfield  {title} {\bibinfo {title} {Thermodynamics of polarized relativistic matter},\ }\bibfield  {journal} {\bibinfo  {journal} {Journal of High Energy Physics}\ }\textbf {\bibinfo {volume} {2016}},\ \href {https://doi.org/10.1007/jhep07(2016)028} {10.1007/jhep07(2016)028} (\bibinfo {year} {2016})\BibitemShut {NoStop}%
\bibitem [{\citenamefont {Noether}(1918)}]{Noether1918}%
  \BibitemOpen
  \bibfield  {author} {\bibinfo {author} {\bibfnamefont {E.}~\bibnamefont {Noether}},\ }\bibfield  {title} {\bibinfo {title} {Invariante variationsprobleme},\ }\href {http://eudml.org/doc/59024} {\bibfield  {journal} {\bibinfo  {journal} {Nachrichten von der Gesellschaft der Wissenschaften zu G{\"o}ttingen, Mathematisch-Physikalisce Klasse}\ }\textbf {\bibinfo {volume} {1918}},\ \bibinfo {pages} {235} (\bibinfo {year} {1918})}\BibitemShut {NoStop}%
\bibitem [{\citenamefont {Noether}(1971)}]{Noether01011971}%
  \BibitemOpen
  \bibfield  {author} {\bibinfo {author} {\bibfnamefont {E.}~\bibnamefont {Noether}},\ }\bibfield  {title} {\bibinfo {title} {Invariant variation problems},\ }\href {https://doi.org/10.1080/00411457108231446} {\bibfield  {journal} {\bibinfo  {journal} {Transport Theory and Statistical Physics}\ }\textbf {\bibinfo {volume} {1}},\ \bibinfo {pages} {186} (\bibinfo {year} {1971})},\ \Eprint {https://arxiv.org/abs/https://doi.org/10.1080/00411457108231446} {https://doi.org/10.1080/00411457108231446} \BibitemShut {NoStop}%
\bibitem [{\citenamefont {Henneaux}\ and\ \citenamefont {Teitelboim}(1992)}]{HenneauxTeitelboim}%
  \BibitemOpen
  \bibfield  {author} {\bibinfo {author} {\bibfnamefont {M.}~\bibnamefont {Henneaux}}\ and\ \bibinfo {author} {\bibfnamefont {C.}~\bibnamefont {Teitelboim}},\ }\href {http://www.jstor.org/stable/j.ctv10crg0r} {\emph {\bibinfo {title} {Quantization of Gauge Systems}}}\ (\bibinfo  {publisher} {Princeton University Press},\ \bibinfo {year} {1992})\BibitemShut {NoStop}%
\bibitem [{\citenamefont {Montenegro}\ \emph {et~al.}(2026)\citenamefont {Montenegro}, \citenamefont {Savioli},\ and\ \citenamefont {Torrieri}}]{Montenegro2026phf}%
  \BibitemOpen
  \bibfield  {author} {\bibinfo {author} {\bibfnamefont {D.}~\bibnamefont {Montenegro}}, \bibinfo {author} {\bibfnamefont {M.~J. P. D.~D.}\ \bibnamefont {Savioli}},\ and\ \bibinfo {author} {\bibfnamefont {G.}~\bibnamefont {Torrieri}},\ }\href@noop {} {\bibinfo {title} {{Gaussian pseudogauge invariant hydrodynamics with spin}}} (\bibinfo {year} {2026}),\ \Eprint {https://arxiv.org/abs/2604.07657} {arXiv:2604.07657 [hep-th]} \BibitemShut {NoStop}%
\bibitem [{\citenamefont {Amoretti}\ \emph {et~al.}(2023)\citenamefont {Amoretti}, \citenamefont {Brattan}, \citenamefont {Martinoia},\ and\ \citenamefont {Matthaiakakis}}]{Amoretti:2022ovc}%
  \BibitemOpen
  \bibfield  {author} {\bibinfo {author} {\bibfnamefont {A.}~\bibnamefont {Amoretti}}, \bibinfo {author} {\bibfnamefont {D.~K.}\ \bibnamefont {Brattan}}, \bibinfo {author} {\bibfnamefont {L.}~\bibnamefont {Martinoia}},\ and\ \bibinfo {author} {\bibfnamefont {I.}~\bibnamefont {Matthaiakakis}},\ }\bibfield  {title} {\bibinfo {title} {{Non-dissipative electrically driven fluids}},\ }\href {https://doi.org/10.1007/JHEP05(2023)218} {\bibfield  {journal} {\bibinfo  {journal} {JHEP}\ }\textbf {\bibinfo {volume} {05}},\ \bibinfo {pages} {218}},\ \Eprint {https://arxiv.org/abs/2211.05791} {arXiv:2211.05791 [hep-th]} \BibitemShut {NoStop}%
\bibitem [{\citenamefont {Amoretti}\ \emph {et~al.}(2024)\citenamefont {Amoretti}, \citenamefont {Brattan}, \citenamefont {Martinoia},\ and\ \citenamefont {Rongen}}]{Amoretti:2024jig}%
  \BibitemOpen
  \bibfield  {author} {\bibinfo {author} {\bibfnamefont {A.}~\bibnamefont {Amoretti}}, \bibinfo {author} {\bibfnamefont {D.~K.}\ \bibnamefont {Brattan}}, \bibinfo {author} {\bibfnamefont {L.}~\bibnamefont {Martinoia}},\ and\ \bibinfo {author} {\bibfnamefont {J.}~\bibnamefont {Rongen}},\ }\bibfield  {title} {\bibinfo {title} {{Dissipative electrically driven fluids}},\ }\href {https://doi.org/10.1007/JHEP12(2024)114} {\bibfield  {journal} {\bibinfo  {journal} {JHEP}\ }\textbf {\bibinfo {volume} {12}},\ \bibinfo {pages} {114}},\ \Eprint {https://arxiv.org/abs/2407.18856} {arXiv:2407.18856 [cond-mat.stat-mech]} \BibitemShut {NoStop}%
\bibitem [{\citenamefont {Kovtun}\ and\ \citenamefont {Shukla}(2020)}]{Kovtun_2020}%
  \BibitemOpen
  \bibfield  {author} {\bibinfo {author} {\bibfnamefont {P.}~\bibnamefont {Kovtun}}\ and\ \bibinfo {author} {\bibfnamefont {A.}~\bibnamefont {Shukla}},\ }\bibfield  {title} {\bibinfo {title} {Einstein’s equations in matter},\ }\bibfield  {journal} {\bibinfo  {journal} {Physical Review D}\ }\textbf {\bibinfo {volume} {101}},\ \href {https://doi.org/10.1103/physrevd.101.104051} {10.1103/physrevd.101.104051} (\bibinfo {year} {2020})\BibitemShut {NoStop}%
\bibitem [{\citenamefont {Weinberg}(2013)}]{Weinberg:1996kr}%
  \BibitemOpen
  \bibfield  {author} {\bibinfo {author} {\bibfnamefont {S.}~\bibnamefont {Weinberg}},\ }\href {https://doi.org/10.1017/CBO9781139644174} {\emph {\bibinfo {title} {{The quantum theory of fields. Vol. 2: Modern applications}}}}\ (\bibinfo  {publisher} {Cambridge University Press},\ \bibinfo {year} {2013})\BibitemShut {NoStop}%
\bibitem [{\citenamefont {Jonsson}\ and\ \citenamefont {Kim}(2026)}]{jonsson2026hydrodynamicscospansfieldtheories}%
  \BibitemOpen
  \bibfield  {author} {\bibinfo {author} {\bibfnamefont {S.}~\bibnamefont {Jonsson}}\ and\ \bibinfo {author} {\bibfnamefont {H.}~\bibnamefont {Kim}},\ }\href {https://arxiv.org/abs/2603.09013} {\bibinfo {title} {Hydrodynamics as cospans of field theories into the bf theory}} (\bibinfo {year} {2026}),\ \Eprint {https://arxiv.org/abs/2603.09013} {arXiv:2603.09013 [hep-th]} \BibitemShut {NoStop}%
\bibitem [{\citenamefont {Lee}\ and\ \citenamefont {Wald}(1990)}]{Lee:1990nz}%
  \BibitemOpen
  \bibfield  {author} {\bibinfo {author} {\bibfnamefont {J.}~\bibnamefont {Lee}}\ and\ \bibinfo {author} {\bibfnamefont {R.~M.}\ \bibnamefont {Wald}},\ }\bibfield  {title} {\bibinfo {title} {{Local symmetries and constraints}},\ }\href {https://doi.org/10.1063/1.528801} {\bibfield  {journal} {\bibinfo  {journal} {J. Math. Phys.}\ }\textbf {\bibinfo {volume} {31}},\ \bibinfo {pages} {725} (\bibinfo {year} {1990})}\BibitemShut {NoStop}%
\bibitem [{\citenamefont {Wald}\ and\ \citenamefont {Zoupas}(2000)}]{Wald:1999wa}%
  \BibitemOpen
  \bibfield  {author} {\bibinfo {author} {\bibfnamefont {R.~M.}\ \bibnamefont {Wald}}\ and\ \bibinfo {author} {\bibfnamefont {A.}~\bibnamefont {Zoupas}},\ }\bibfield  {title} {\bibinfo {title} {{A General definition of 'conserved quantities' in general relativity and other theories of gravity}},\ }\href {https://doi.org/10.1103/PhysRevD.61.084027} {\bibfield  {journal} {\bibinfo  {journal} {Phys. Rev. D}\ }\textbf {\bibinfo {volume} {61}},\ \bibinfo {pages} {084027} (\bibinfo {year} {2000})},\ \Eprint {https://arxiv.org/abs/gr-qc/9911095} {arXiv:gr-qc/9911095} \BibitemShut {NoStop}%
\bibitem [{\citenamefont {Harlow}\ and\ \citenamefont {Wu}(2020)}]{Harlow_2020}%
  \BibitemOpen
  \bibfield  {author} {\bibinfo {author} {\bibfnamefont {D.}~\bibnamefont {Harlow}}\ and\ \bibinfo {author} {\bibfnamefont {J.-q.}\ \bibnamefont {Wu}},\ }\bibfield  {title} {\bibinfo {title} {Covariant phase space with boundaries},\ }\bibfield  {journal} {\bibinfo  {journal} {Journal of High Energy Physics}\ }\textbf {\bibinfo {volume} {2020}},\ \href {https://doi.org/10.1007/jhep10(2020)146} {10.1007/jhep10(2020)146} (\bibinfo {year} {2020})\BibitemShut {NoStop}%
\bibitem [{\citenamefont {Gordon}(1928)}]{Gordon1928}%
  \BibitemOpen
  \bibfield  {author} {\bibinfo {author} {\bibfnamefont {W.}~\bibnamefont {Gordon}},\ }\bibfield  {title} {\bibinfo {title} {Der strom der diracschen elektronentheorie},\ }\href {https://doi.org/10.1007/BF01327881} {\bibfield  {journal} {\bibinfo  {journal} {Zeitschrift f{\"u}r Physik}\ }\textbf {\bibinfo {volume} {50}},\ \bibinfo {pages} {630} (\bibinfo {year} {1928})}\BibitemShut {NoStop}%
\end{thebibliography}%

\setcounter{equation}{0}
\onecolumngrid
\appendix

\section{Computation of the Belinfante-Rosenfeld algebra}\label{AppA}

In this section, I give a detailed derivation on the Belinfante-Rosenfeld algebras of Minkowski and Einstein-Cartan spacetimes mentioned in the main text. 

\subsection{Minkowski space}

The Belinfante-Rosenfeld transformation in Minkowski spacetime is given by
\begin{equation}
    \label{EqApp:BRM}
    \delta\eta = d\xi ~,~ \delta B^I = D\Xi^I +\xi C^I_JB^J~,
\end{equation}
with all operators defined around Eq.\eqref{EqMink:ConBi3} of the main text.

\noindent Assume two transformations parametrized by $(\xi_1,\Xi^I_1)$ and $(\xi_2,\Xi^I_2)$.
\begin{align}
\label{EqApp:Baction}
    \delta_1\delta_2B^I &= \delta_1\left(D\Xi^I_2 + \xi_2 C^I_JB^J\right) = \delta_1(-\eta C^I_J\Xi^J_2) + \xi_2 C^I_J\delta_1 B^J = -d\xi_1C^I_J \Xi^J_2 + \xi_2C^I_J\left(D\Xi^J_1 + \xi_1 C^J_KB^K\right) \nonumber
    \\
    & = - d(\xi_1C^I_J\Xi^J_2) + \xi_1C^I_Jd\Xi^J_2 + \xi_2 C^I_J d\Xi^J_1 = -D(\xi_1C^I_J\Xi^J_2)+ \xi_1C^I_Jd\Xi^J_2 + \xi_2 C^I_J d\Xi^J_1~,
\end{align}
and $\delta_1\delta_2\eta =0$. In the first line of \eqref{EqApp:Baction}, I employed the linearity of $\delta$ and the fact the BR parameters do not transform. In the second line, I used the Leibnitz rule satisfied by $d$ and made liberal use of $C^I_JC^J_K = 0$.

\noindent Therefore, the Belinfante-Rosenfeld algebra in Minkowski space reads
\begin{equation}
\label{EqApp:MinkAlgebra}
    [\delta_1, \delta_2]B^I = (\delta_1\delta_2 - \delta_2\delta_1)B^I = D\left(\xi_2C^I_J\Xi^J_1-\xi_1C^I_J\Xi^J_2\right) \equiv D\Xi^I_3 = \delta_3 B^I~,
\end{equation}
with the parameters of $\delta_3$ as given in \eqref{EqMink:ThirdT}.

\subsection{Einstein-Cartan spacetime}

The Belinfante-Rosenfeld transformation in an Einstein-Cartan spacetime is given in \eqref{EqEC:BR2} of the main text. Namely,
\begin{align}
    &\delta\eta = \eta\cdot\theta ~,~\delta\omega = \nabla\theta~,~\delta B^2 = B^2\cdot\theta+ \nabla\Xi^2~,~  \nonumber
    \\
    \label{EqApp:BR2}
    \\
    &\delta B^1= B^1\cdot\theta +\nabla\Xi^1 -(\iota_\eta\tau)\cdot_R\Xi^1 +(\iota_\eta\Omega)\cdot_R\Xi^2~.\nonumber
\end{align}
As in Minkowski spacetime, assume two transformations parametrized by $(\theta_1,\Xi^I_1)$ and $(\theta_2,\Xi^I_2)$. Let us calculate the algebra of \eqref{EqApp:BR2} on each field at increasing levels of difficulty. Begin with $\eta$,
\begin{equation}
    \delta_1\delta_2\eta = \delta_1\eta\cdot\theta_2 = (\eta\cdot\theta_1)\cdot\theta_2 = \eta\cdot(\theta_1\cdot\theta_2) \Rightarrow [\delta_1,\delta_2]\eta = \eta\cdot[\theta_1,\theta_2]~,
\end{equation}
where I used the associativity of the $\cdot$ matrix product and the fact $(\theta, \Xi^I)$ do not transform under the $\delta$s.

Next consider the connection $\omega$
\begin{equation}
    \delta_1\delta_2\omega = \delta_1(\nabla\theta_2) = \delta_1[\omega,\theta_2] = [\delta_1\omega, \theta_2] = [\nabla\theta_1, \theta_2] \Rightarrow [\delta_1,\delta_2]\omega = [\nabla\theta_1, \theta_2]-[\nabla\theta_2, \theta_1] = \nabla[\theta_1,\theta_2],
\end{equation}
where I used the linearity of $\delta$, the anti-commutativity of the graded commutator for $L$ $0$-forms and the Leibnitz rule satisfied by $\nabla$ with respect to the graded commutator (can be verified directly via Eq.\eqref{EqEC:GradedCom} in the main text).

Up next, consider $B^2$,
\begin{align}
    \delta_1\delta_2B^2 &= \delta_1\left(B^2\cdot\theta_2 + \nabla\Xi^2_2\right) = \left(B^2\cdot\theta_1 + \nabla\Xi^2_1\right)\cdot\theta_2 + \delta_1[\omega, \Xi^2_2] -\delta_1\eta\Xi^1_2 \nonumber
    \\
    &=B^2\cdot(\theta_1\cdot\theta_2) + \nabla\Xi^2_1\cdot\theta_2 + [\nabla\theta_1, \Xi^2_2] - (\eta\cdot\theta_1)\Xi^1_2 \nonumber
    \\
    & = B^2\cdot(\theta_1\cdot\theta_2) + \nabla(\Xi^2_1\cdot\theta_2) + [\nabla\theta_2,\Xi^2_1] + [\nabla\theta_1,\Xi^2_2]-(\eta\cdot\theta_1)\Xi^1_2 -(\eta\cdot \theta_2)\Xi^1_1~.
\end{align}
In the first and second lines, I have used again the linearity of $\delta$ and \eqref{EqApp:BR2}. In the last, I collected and added terms to produce a total derivative, and arranged the remainder so it becomes obvious it is symmetric under the interchange of their lower $1,2$ indices. Therefore, the commutator on $B^2$ closes to
\begin{equation}
    [\delta_1,\delta_2]B^2 = B^2\cdot[\theta_1,\theta_2] + \nabla\left(\Xi^2_1\cdot\theta_2 - \nabla\Xi^2_2\cdot\theta_1\right) \equiv B^2\cdot\theta_3 + \nabla\Xi^2_3 = \delta_3 B^2~,
\end{equation}
with obvious definitions for the parameters of $\delta_3$. We already see the semi-direct product nature of the Belinfante-Rosenfeld group in Einstein-Cartan spacetime, since $\Xi^2_3$ depends both on the $\theta$s and $\Xi^I$s. 

Let us now complete our calculation with the algebra action on $B^1$. To aid us in that, we need first two auxiliary formulas regarding the transformation properties of the torsion and curvature bi-forms. We have
\begin{equation}
    \delta\tau = \delta\nabla\eta = \nabla\delta \eta +[\delta\omega,\eta] = (\nabla\eta)\cdot\theta - \eta\cdot\nabla\theta +[\nabla\theta,\eta] = \tau\cdot\theta~,
\end{equation}
and
\begin{equation}
\delta\Omega = \delta\left(d\omega + {1\over 2}[\omega,\omega]\right) = d\delta\omega + [\omega,\delta\omega] = \nabla\delta\omega = [\Omega,\theta]= \Omega\cdot\theta~.    
\end{equation}
Thus, I have shown that $\tau$ and $\Omega$ transform as $R$-tensors under $R$-rotations. This holds even for their contracted variants $\iota_\eta\tau$ and $\iota_\eta\Omega$, since $\eta$ also transforms as an $R-$tensor under rotations. This is in hindsight obvious, since the energy-momentum conservation equation is rotationally invariant and the only free $R$-indices on the right-hand side of said equation are the ones due to $\iota_\eta$.

Finally, we have for $\delta_1\delta_2B^1$,
\begin{align}
    \delta_1\delta_2B^1 &= \delta_1\left(B^1\cdot\theta_2 + \nabla\Xi^1_2 -(\iota_\eta\tau)\cdot_R\Xi^1_2 +(\iota_\eta\Omega)\cdot_R\Xi^2_2\right) 
    \nonumber
    \\
    &=\left(B^1\cdot\theta_1 + \nabla\Xi^1_1 -(\iota_\eta\tau)\cdot_R\Xi^1_1 +(\iota_\eta\Omega)\cdot_R\Xi^2_1\right)\cdot\theta_2 +[\nabla\theta_1,\Xi^1_2]-[(\iota_\eta\tau)\cdot_R\Xi^1_2]\cdot\theta_1 + [(\iota_\eta\Omega)\cdot_R\Xi^2_2]\cdot\theta_1
    \nonumber
    \\
    &= \left(B^1\cdot\theta_1 + \nabla\Xi^1_1\right)\cdot\theta_2 + [\nabla\theta_1,\Xi^1_2] -[(\iota_\eta\tau)\cdot_R\Xi^1_1]\cdot\theta_2 +[(\iota_\eta\Omega)\cdot_R\Xi^2_1]\cdot\theta_2-[(\iota_\eta\tau)\cdot_R\Xi^1_2]\cdot\theta_1 + [(\iota_\eta\Omega)\cdot_R\Xi^2_2]\cdot\theta_1
    \nonumber
    \\
    &= B^1\cdot(\theta_1\cdot\theta_2) +\nabla(\Xi^1_1\cdot\theta_2) + {\rm symmetric}~.
\end{align}
``Symmetric" means the last equality is written up to terms dropping from the final commutator. Hence, the algebra on $B^1$ closes to 
\begin{equation}
    [\delta_1,\delta_2]B^1 = \delta_3 B^1,
\end{equation}
with the parameters of $\delta_3$ being
\begin{equation}
\label{EqApp:AlgebraFinal}
    (\theta_3, \Xi^I_3) = ([\theta_1,\theta_2], \Xi^I_1\cdot\theta_2-\Xi^I_2\cdot\theta_1)~.
\end{equation}
Equation \eqref{EqApp:AlgebraFinal} is our final result for the algebra of the Belinfante-Rosenfeld group in Einstein-Cartan spacetime and showcases the group structure advertised in Eq.\eqref{EqEC:BRtGroup} of the main text. 

\section{Examples}\label{AppB:Examples}

In this appendix, I present examples on how to apply the formalism in the main text to recover well-known results. We also define a new choice of BRt that symmetrizes the energy-momentum tensor that differs from the Belinfante-Rosenfeld one \cite{BELINFANTE1940449,Rosenfeld1940}. We split this appendix into two sections, section \ref{AppBSec:EMS} discusses energy-momentum and angular momentum conservation, while section \ref{AppBSec:Charge} energy-momentum and U(1)-charge conservation.

\subsection{Conserved energy-momentum and angular momentum}\label{AppB:EMJExamples}\label{AppBSec:EMS}

Let us consider energy-momentum and angular momentum conservation in Minkowski space and their corresponding BRt. We repeat the relevant equations below for ease of reference,

\begin{equation}
\label{AppBEq:EMJCon}
    {\cal F}^I = DB^I = 0~,
\end{equation}
\begin{equation}
\label{AppBEq:BRtMin}
    \delta \eta = d\xi ~,~ \delta B^I = D\Xi^I +\xi C^I_JB^J~,~\delta {\cal F}^I = \xi C^I_J{\cal F}^J - \tau C^I_J\Xi^J.
\end{equation}

We may use the BRt \eqref{AppBEq:BRtMin} to fix particular forms of $B^I$. A popular choice is the Belinfante-Rosenfeld one, for which the energy-momentum tensor is symmetric viz. $\eta B^1_{\rm BR} = 0$ \cite{BELINFANTE1940449, Rosenfeld1940}. To enforce this choice we must fix $\Xi^1 = \Xi^1_{\rm BR}$ such that
\begin{align}
\label{AppBEq:BRgauge}
    &\eta(B^1 + d\Xi^1_{\rm BR}) = 0 \Leftrightarrow \eta d\Xi^1_{\rm BR} = - \eta B^1 = -dB^2
    \nonumber
    \\
   \Leftrightarrow~ &\eta d\Xi^I_{\rm BR} = d(\eta\Xi^1_{\rm BR}) = - d B^2\Leftrightarrow \eta\Xi^1_{\rm BR} = - B^2 + d\Xi^2_{\rm res}~. 
\end{align}
To arrive at \eqref{AppBEq:BRgauge}, we used the conservation equations and $\tau = 0$. The final result tells us that to symmetrize the energy-momentum tensor, we must fix the \textit{anti-symmetric} part of $\Xi^1$ to equal the spin tensor, up to a residual transformation of $B^2$ with $\xi = 0,~ \Xi^2 = \Xi^2_{\rm res}$. It is easy then to see that the transformed spin tensor reads $B^2_{\rm BR} = d\Xi^2_{\rm res}$. If no boundary spin currents are present, we may set $\Xi^2_{\rm BR} = 0$ and $B^2_{\rm BR} = 0$. This is the well-known result found in the literature, see e.g. \cite{Speranza_2021}, expressed in terms of biforms.

Instead of symmetrizing $B^1$, we may choose $(\xi, \Xi^I)$ to set the spin tensor to zero. Because of angular momentum conservation, $B^1$ will necessarily be symmetric when the spin vanishes. To set $B^2 = 0$ after a BRt, we require
\begin{equation}
\label{AppBEq:SpinCon}
\delta B^2 = - B^2 \Leftrightarrow d\Xi^2 - \eta\Xi^1 + \xi B^1 = -B^2~.
\end{equation}
There are three independent ways to satisfy \eqref{AppBEq:SpinCon}. The first is to set
\begin{equation}
    \eta \Xi^1 = \eta\Xi^1_{\rm BR} = -B^2~,~\xi = 0 = \Xi^2.
\end{equation}
This is a slightly more constrained version of the Belinfante-Rosenfeld choice \eqref{AppBEq:BRgauge}. 

The second approach at $B^2 = 0$ requires $\Xi^1 = 0 = \xi$ and
\begin{equation}
\label{AppBEq:SpinTop}
    d\Xi^2 = - B^2~.
\end{equation}
The existence of a $\Xi^2$ satisfying \eqref{AppBEq:SpinTop} is not guaranteed, and requires that either spin is confined on the boundary of our spacetime or that $\Xi^2$ is multiply-valued. BRts of the form \eqref{AppBEq:SpinTop} are also well-known in the literature, see e.g. \cite{HEHL197655}.

Finally, we may set $\Xi^1 = 0 =\Xi^2$ and fix $\xi$ such that
\begin{equation}
\label{AppBEq:SpinCo}
    \xi B^1 = - B^2~. 
\end{equation}
Physically, \eqref{AppBEq:SpinCo} amounts to measuring the total angular momentum from the point $x_R+\xi$, instead of $\xi$.\footnote{Recall $J =\int_{\Sigma_L} x_RB^1 - B^2$.} Using different reference points to measure angular momentum has been discussed in detail in \cite{Speranza_2021}. Our approach differs from the approach in \cite{Speranza_2021}, since the authors fix a BRt with a non-zero $\Xi^1$ and then evaluate what reference point this choice corresponds to. Equation \eqref{AppBEq:SpinCo} instead fixes the BRt and the reference point simultaneously.

\subsection{Conserved energy-momentum and U(1) current}\label{AppBSec:Charge}

The energy-momentum and charge conservation equations and their corresponding BRt are
\begin{align}
     &dB^1 = (\iota_\eta F) j~,~ dj =0~,
     \\
     \label{AppBEq:ChargeBRt}
      &\delta B^1 = d\Xi^1 -\iota_\eta F\Xi^2~,~ \delta j = d\Xi^2~.
\end{align}
The transformation \eqref{AppBEq:ChargeBRt} may also be expressed as
\begin{equation}
\label{AppBEq:BRtCharge}
    \delta B^1 = d(\Xi^1-\iota_\eta A \Xi^2) + \iota_\eta A d\Xi^2~,~\delta j = d\Xi^2~,
\end{equation}
which is the form of the transformation usually employed. 

As a first example, consider Maxwell theory in vacuum. We may symmetrize the energy-momentum tensor of E$\&$M,  while leaving the charge current intact by setting $\Xi^1 = 0$, $\Xi^2 = \star F$. We could also set $\Xi^1 = -A\star F$, $\Xi^2 = 0$  \cite{landau}. 

A second example is that of the Gordon decomposition of the U(1) current of a $4d$ free Dirac fermion \cite{Gordon1928}. In this case, $\Xi^1 = 0$ and 
\begin{equation}
    \Xi^2 \propto \star \bar{\psi}\Sigma^{\mu\nu}\psi~,~\Sigma^{\mu\nu} = {i\over 4}[\gamma^{\mu},\gamma^\nu]~,
\end{equation}
with $ \gamma^\mu$ Dirac's gamma matrices. The Gordon decomposition acts on the energy-momentum tensor by adding a Pauli interaction term $\iota_\eta F \Xi^2$ or in components $F_{ \nu a}\bar{\psi}\Sigma^{\nu}_{~\mu}\psi$. 

\section{Belinfante-Rosenfeld transformation from an action point of view}\label{AppC:ActionBRt}

In this appendix, I consider whether a BRt can be lifted to a symmetry of the action of a system. Since the BRt is a symmetry of the conservation equations when said equations hold, this seems unlikely. To avoid the on-shell nature of the conservation laws and discuss BRts at the level of the action, we may use Noether's second theorem. To this end, we must couple a given theory to background gauge fields. Assume a matter theory with a conserved U(1) current $j$ and an energy-momentum tensor $B^1$.\footnote{This is the simplest setup, but can be extended to any theory.} Their couplings to background fields read
\begin{equation}
\label{Eq1}
    S_{c} =  \int A j + \eta\cdot_R B^1 ~,
\end{equation}
where $A$ is a U(1) gauge field, and $\cdot_R$ and $\eta$ as defined in the main text around Eq.s\eqref{EqEC:BR1} and \eqref{EqMink:RAntisym}, respectively. For $S_c$, the conservation of $j$ and non-conservation of $B^1$ hold off-shell by assuming invariance under gauge transformations of the background fields.\smallskip

\noindent Under the BR transformation in Eq.(44) of the main text, $S_c$ transforms as
\begin{align}
\label{Eq:DeltaS}
    \delta S_c &= \int A d\Xi^2 +\eta\cdot_R(d\Xi^1-\iota_\eta F\Xi^2) \nonumber
    \\
    &= \int F\Xi^2-\eta\cdot_R(\iota_\eta F \Xi^2) +\tau\cdot_R\Xi^1 - d(A \Xi^2- \eta\cdot_R \Xi^1)\nonumber
    \\
    & = -\int d(A \Xi^2- \eta\cdot_R \Xi^1)~.
\end{align}
To arrive at the first equality of \eqref{Eq:DeltaS}, we used Eq.(44) of the main text. In the second equality, we moved the exterior derivatives from $\Xi^1$ and $\Xi^2$ to $A$ and $\eta$. Finally, we used the equalities $\eta\cdot_R(\iota_\eta F \Xi^2) = F \Xi^2$ and $\tau = 0$. 

\noindent We see that $S_c$ changes by a total boundary term, so the BR transformation is a symmetry of $S_c$. This is one point of view of the BR transformation. 

However the point of view advocated in the paper, i.e. BR transformations shift charges from the matter to the gauge sector or vice versa, is one that incorporates dynamical gauge fields into the above picture. So let us assume $A$ and $\eta$ are dynamical with an action
\begin{equation}
    S_{\rm gauge} = -\int FH^2 + \tau\cdot_R H^1~,
\end{equation}
where $H^1$, $H^2$ are the biforms in Eq. \eqref{EqEC:Maxwell} of the main text. The action $S_{\rm gauge}$ can be thought of as describing electromagnetism and teleparallel gravity in matter. 

\noindent Using integration by parts, we may express $S_{\rm gauge}$ as 
\begin{equation}
    S_{\rm gauge} = -\int AdH^2 + \eta\cdot_R dH^1 + d(AH^2 + \eta\cdot_RH^1).
\end{equation}
Thus the total action reads
\begin{equation}
\label{Eq:Bothactions}
    S_{\rm tot} = S_c + S_{\rm gauge} = \int A(-dH^2+j) + \eta\cdot_R(-dH^1 + B^1) -d(AH^2+ \eta\cdot_RH^1)~.
\end{equation}
The form of $S_{\rm tot}$ implies that a change in $S_c$ due to a BR transformation, Eq.\eqref{Eq:DeltaS}, can be compensated by a change in $H^1$ and $H^2$ as long as the charges defined by the fluxes of $H^1$ and $H^2$ change to include $\Xi^1$ and $\Xi^2$. This is the main conclusion of the paper, derived at the level of an action.

Note that the analysis above works only because the actions we used can be written in a form linear in the background fields $A$ and $\eta$. In general, the BR transformation of the currents can be compensated by a shift of the first variation of $S$ with respect to $F$ and $\tau$. For the simplest possible case, we may set $H^2 = \star F$ and $H^1 = \star \tau$. Requiring the transformations of $H^1$ and $H^2$ compensate the current BRt, we are led to Eq.\eqref{Eq:Conditions} examined in the main text.

 \end{document}